\documentclass[onecolumn,useAMS,usenatbib]{mn2e}
\usepackage{graphicx,natbib}
\usepackage{url}
\usepackage{natbib}

\usepackage{bm}
\usepackage{amsmath}
\usepackage{amssymb}
\usepackage{epsf}

\usepackage{color} 
\newcommand{\bea}{\begin{eqnarray}}
\newcommand{\eea}{\end{eqnarray}}
\newcommand{\rvir}{r_{\rm vir}}
\newcommand{\rpara}{r_{\parallel}}
\newcommand{\rperp}{r_{\perp}}

\newcommand{\thetavir}{\theta_{\rm vir}}
\newcommand{\Mvir}{M_{\rm vir}}
\newcommand{\Deltavir}{\Delta_{\rm vir}}
\newcommand{\ek}{e_{\kappa}}
\newcommand{\ephi}{e_{\varphi}}
\newcommand{\ekappa}{e_{\kappa}}

\newcommand{\KB}{k_{\rm B}}
\newcommand{\TCMB}{T_{\rm CMB}}
\newcommand{\ICMB}{I_{\rm CMB}}
\newcommand{\Mgas}{M_{\rm gas}}
\newcommand{\Tgas}{T_{\rm gas}}
\newcommand{\fgas}{f_{\rm gas}}

\newcommand{\barS}{\overline{S}}
\newcommand{\barSvar}{\overline{S^2}}

\newcommand{\Sobs}{S_{\rm obs}}
\newcommand{\nobs}{n_{\rm obs}}
\newcommand{\Nobs}{N_{\rm obs}}

\newcommand{\Omegaobs}{\Omega_{\rm obs}}

\newcommand{\Msun}{M_{\odot}}

\newcommand{\simgt}{\lower.5ex\hbox{$\; \buildrel > \over \sim \;$}}
\newcommand{\simlt}{\lower.5ex\hbox{$\; \buildrel < \over \sim \;$}}

\title{Lensing Magnification: Implications for Counts of Submillimeter Galaxies and SZ Clusters}

\author[Lima, Jain and Devlin]{
Marcos Lima\thanks{{\tt mlima@sas.upenn.edu}},
Bhuvnesh Jain and
Mark Devlin
\\
Department of Physics \& Astronomy, University of Pennsylvania, Philadelphia, PA 19104 \\
}

\date{\today} 

\begin{document}
\maketitle

\begin{abstract}
We study lensing magnification of source galaxies by intervening galaxy groups and clusters using a halo model. Halos are modeled with truncated NFW profiles with ellipticity added to
their lensing potential and propagated to observable lensing statistics.
We present the formalism to calculate observable effects due to a distribution of halos of different masses at different redshifts along the line of sight.  We calculate the effects of magnification on the number counts of
high-redshift galaxies. Using BLAST survey data for submillimeter galaxies (SMGs), we find that magnification affects the steep, high flux part of the counts by about 60\%.
The effect becomes much stronger if the intrinsic distribution is significantly steeper
than observed. We also consider the effect of this high-redshift galaxy population on
contaminating the Sunyaev-Zel¡Çdovich (SZ) signal of massive clusters using the halo model approach. We find that for the majority of clusters expected to be detected with ongoing SZ surveys, there is significant contamination from the Poisson noise due to background SMGs. This contribution can be comparable to the SZ increment for typical clusters and can also contaminate the SZ decrement of low
mass clusters. Thus SZ observations, especially for the increment part
of the SZ spectrum, need to include careful modeling of this
irreducible contamination for mass estimation. Lensing further
enhances the contamination, especially close to the cores of massive
clusters and for very disturbed clusters with large magnification cross-section. 
\end{abstract}

\vspace{-0.3in}

\section{Introduction}

Gravitational lensing, the deflection of light rays from background galaxies
by intervening objects, has a number of cosmological and astrophysical 
applications
\citep{SchEhlFal92,BlaNar92,NarBar96,BarSch01,SchKocWam06,Hoejai08}.  
In the weak lensing regime, small distortions
of source galaxies can statistically constrain cosmological parameters 
related to structure formation and theories of gravity. 
In the strong lensing regime, where light is
deflected by rare massive galaxy clusters, it is also possible to 
infer detailed information on cluster profiles, though with less statistics. 
Although lensing conserves surface brightness
of sources, it changes their observed fluxes and sizes; e.g. background 
galaxies have their fluxes magnified in the line-of-sight of massive 
halos cores.    

Galaxy clusters that produce dramatic lensing effects are themselves
potentially powerful cosmological probes 
\citep{WanSte98, HaiMohHol01}, provided one can detect them
in pure samples, and measure their masses and redshifts precisely
\citep{Hu03,LimHu05,LimHu07, Rozetal07,Rozetal09}. 
Techniques for cluster
detection and mass measurement include the counting of optical galaxies, 
the measurement of their lensing signal, their X-ray temperature or flux and 
their Sunyaev-Zel'dovich flux decrement/increment (SZ effect). 
The SZ effect \citep{SunZel72,Ito98,Bir99,CarHolRee02} 
results from the up-scatter of cosmic microwave background (CMB) 
photons by the hot
electrons in the intra-cluster medium, which shifts the underlying spectrum
and causes a decrement/increment at wavelengths longer/shorter than $\sim$ 1 mm. 
This allows for the detection of clusters independently of their
redshifts and the measurement of their masses, as long as the effects of possible 
contaminants, such as radio sources and far-IR/submillimeter galaxies can be 
removed or at least well understood.

The submillimeter galaxy population (SMGs) consists of dusty high-redshift 
galaxies with high star formation rates 
\citep{BlaLon93,Baretal98,Hughes98,Blaetal02}. 
The energy output of the star forming regions 
heats up dust grains, which emit a modified blackbody 
spectrum that peaks around 0.1 mm in the rest-frame. 
SMGs can produce a non-negligible point-source contribution to CMB anisotropies 
\citep{ScoWhi99}.
When positioned on the line-of-sight of cluster cores, these galaxies can 
contaminate the cluster SZ flux, filling up the decrement and enhancing
the increment, resulting in mis-estimation of the cluster mass from 
SZ measurements
\citep{KnoHolChu04, WhiMaj04}. In addition, galaxy clusters can magnify
the fluxes of the SMGs
\citep{Bla96, Bla97, Bla99, PacScoCha08}, 
changing their counts distribution and further enhancing 
the contamination of the SZ signal.

In this paper, we study the effect of lensing magnification of SMGs 
by foreground clusters from a twofold perspective. On the one
hand, galaxy clusters magnify the fluxes of SMGs, changing their intrinsic
counts distribution, but potentially facilitating
their detection and follow-up studies. 
On the other hand, the magnified fluxes of SMGs can contaminate unresolved SZ clusters 
and may need to be understood in order to provide clean SZ mass estimates. 
For instance, \cite{DiePar09} find that including point sources in the modelling
of cluster masses results in better agreement between x-ray and SZ mass
estimates.
Follow-up observations of unresolved clusters provide a means to tackle these 
issues, allowing for separation and removal of SZ contaminants and detailed
studies of their properties.

We start in \S~\ref{sec:halo} describing the properties of the dark matter halos
used to represent the lens galaxy clusters. 
In \S~\ref{sec:lensing} we briefly 
review the basic formalism to analytically compute lensing magnification by 
massive halos and in \S~\ref{sec:prob} we describe the lensing probability and 
some of its applications. 
Our main results are shown in \S~\ref{sec:results}, where we estimate 
the effects of lensing on the SMGs distributions (\S~\ref{subsec:SMG})
and show the contamination effect of the (lensed) SMG population 
on the SZ flux of clusters (\S~\ref{subsec:SZ}). 
We conclude in \S~\ref{sec:conclusion}.  

Throughout, we assume a fiducial cosmology for a flat universe with parameter 
values based on the results of the Wilkinson Microwave Anisotropy Probe 
first year data release (WMAP1) \citep{Spergel2003} 
The cosmological parameters (and their values) 
are the normalization of the initial curvature spectrum 
$\delta_\zeta (=5.07\times 10^{-5})$ at $k=0.05$ Mpc$^{-1}$
(corresponding to $\sigma_8=0.91$), its tilt
$n (=1)$, the baryon density relative to critical
$\Omega_bh^2 (=0.024)$, the matter density
$\Omega_{\rm m} h^2 (=0.14)$, and two dark energy parameters: its density
$\Omega_{\rm DE} (=0.73)$ and equation of state
$w(=-1)$, which we assume to be constant.

\section{Halo Properties} \label{sec:halo}

In this section we present the halo properties and scaling relations
assumed throughout to estimate lensing effects of galaxy clusters. 
We take the NFW profile prescription for dark matter halos \citep{NFW97}
\begin{eqnarray}
\rho(r)=\frac{\rho_s}{(cr/r_{\rm vir})(1+cr/r_{\rm vir})^2}\,,
\end{eqnarray}
where $\rvir$ is the virial radius and the halo concentration $c$ is given by a 
fit to simulations from \citet{Bullock01}
\begin{eqnarray}
c(\Mvir,z)&=&\frac{9}{1+z}\left(\frac{\Mvir}{M_*}\right)^{-0.13}\,,
\end{eqnarray}
with $M_*$ such that $\sigma(M_*)=\delta_c$. Here $\sigma^2(M)$ is the variance of the
linear density field, defined in Eq.~(\ref{eq:sigmaR}) below, and $\delta_c$ is the linearly extrapolated 
density contrast threshold in spherical collapse. Even though $\delta_c$ has a small redshift
and cosmology dependency, here we take it to be fixed at its value in a $\Omega_{\rm m}=1$ universe, 
i.e. $\delta_c=1.686$.
The halo virial mass $\Mvir$ is
\begin{eqnarray}
\Mvir=\int_{0}^{\rvir} dr 4\pi r^2 \rho(r) = \frac{4 \pi\rho_s \rvir^3 }{c^3}f^{-1}\,,
\end{eqnarray}
where $f=(\ln(1+c)-c/(1+c))^{-1}$.
The virial radius $\rvir$ can be computed from the virial overdensity $\Delta_c$ relative 
to {\it critical} density
\begin{eqnarray}
\Delta_c&=&\frac{3\Mvir}{\rho_{\rm crit}(z) 4\pi \rvir^3}\,,
\end{eqnarray}
where $\rho_{\rm crit}(z)=\rho_{\rm crit,0}E^2(z)$ is the critical density at redshift $z$,
$\rho_{\rm crit,0}$ is its value at $z=0$
and the scaled Hubble parameter is 
\begin{eqnarray}
E^2(z)=\frac{H^2(z)}{H_0^2}=\Omega_{\rm m}(1+z)^3+\Omega_{\rm DE}(1+z)^{3(1+w)}\,.
\end{eqnarray}
We take $\Delta_c$ from a fit to simulations
of \citet{BryNor98} for flat cosmologies
\begin{eqnarray}
\Delta_c=18\pi^2+82x-39x^2\,,
\end{eqnarray}
where $x=\omega_{\rm m}(z)-1$ and 
$\omega_{\rm m}(z)=\Omega_{\rm m}(1+z)^3/E^2(z)$.
One can similarly define the overdensity relative to the {\it mean matter} density 
$\Deltavir=\Delta_c/\omega_{\rm m}(z)$.
Finally for the halo distribution in mass and redshift we take 
the description of \citet{SheTor99} for the {\it comoving} differential 
number density of halos per logarithmic mass interval
\begin{eqnarray}
\frac{d n}{d\ln M_{\rm vir}} = {\bar \rho_{\rm m} \over M_{\rm vir}} f(\nu) {d\nu \over d\ln M_{\rm vir}}\,, 
         \label{eqn:massfn}
\end{eqnarray}
where $\nu = \delta_c/\sigma(M_{\rm vir})$ and 
\begin{eqnarray}
\nu f(\nu) = A\sqrt{{2 \over \pi} a\nu^2 } [1+(a\nu^2)^{-p}] \exp[-a\nu^2/2]\,.
\end{eqnarray}
Here
$\sigma^2(M)$ is the variance of the linear density field in a top hat of radius $r$
that encloses $M=4\pi r^3 \bar \rho_{\rm m}/3$ at the background density
\begin{eqnarray}
\sigma^2(r) = \int \frac{d^3k}{(2\pi)^3} |\tilde{W}(kr)|^2 P_{\rm L}(k)\,,
\label{eq:sigmaR}
\end{eqnarray}
where $P_{\rm L}(k)$ is the linear power spectrum and $\tilde W$ is the Fourier transform
of the top hat window. The normalization constant $A$
is such that $\int d\nu f(\nu)=1$ and we take the parameter values $p=0.3$, $a=0.75$. 

\section{Lensing Magnification by Massive Halos} \label{sec:lensing}

\subsection{Axially Symmetric Lenses}

\begin{figure*}
  \resizebox{85mm}{!}{\includegraphics[angle=0]{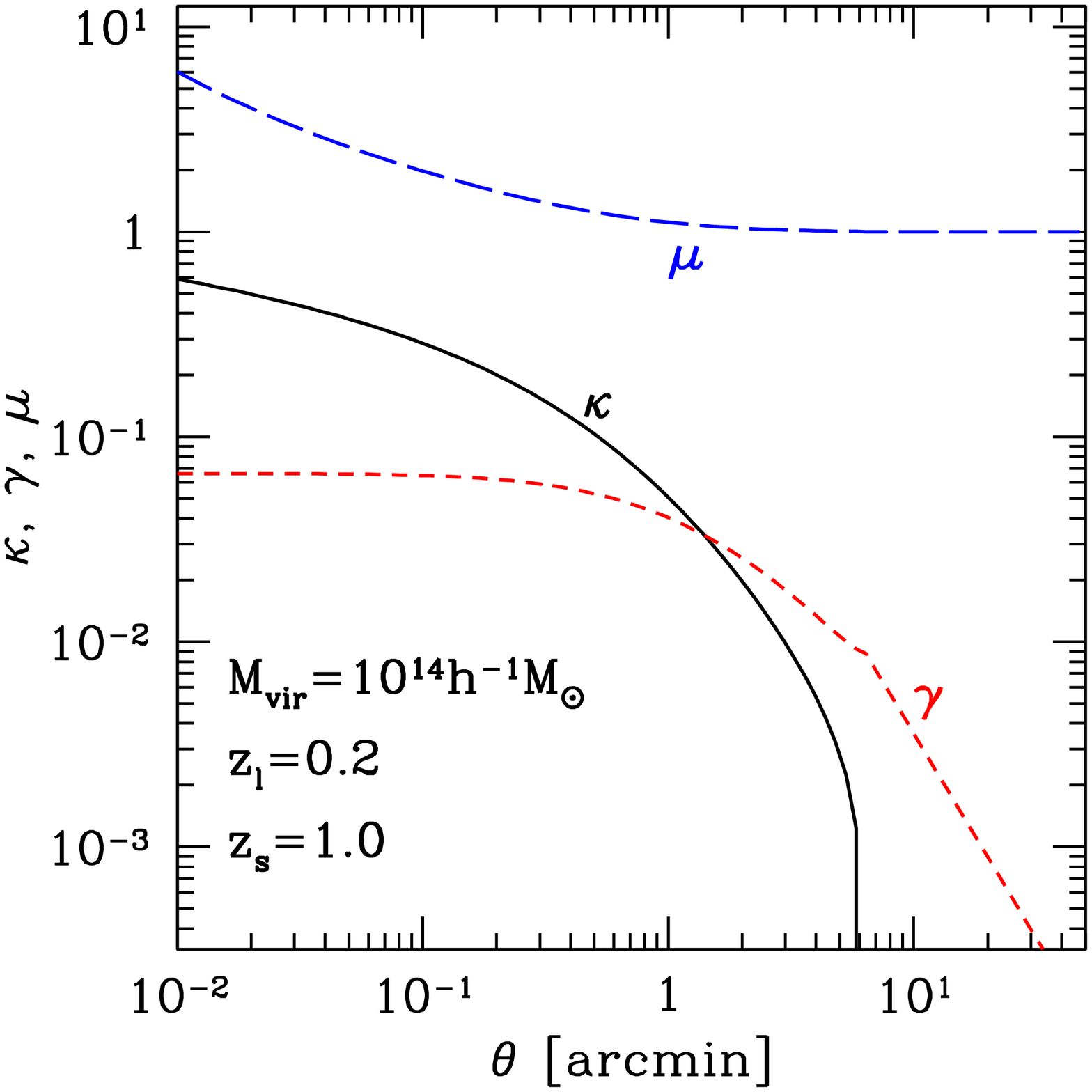}}
  \resizebox{85mm}{!}{\includegraphics[angle=0]{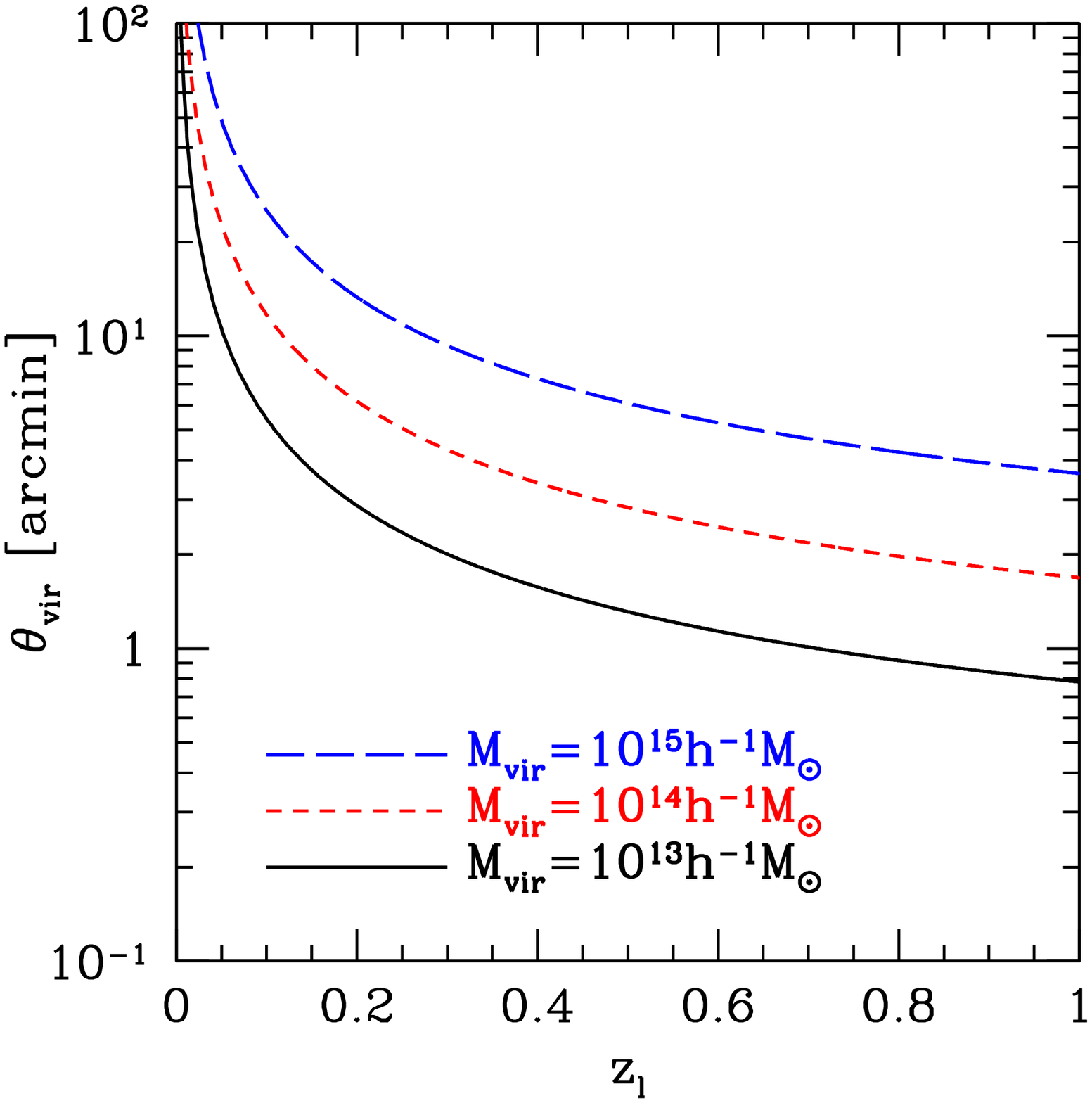}}
  \caption{({\it Left}): Convergence $\kappa$, shear $\gamma$ and magnification 
$\mu$ as a function of angular separation $\theta$ from the halo center. 
Here the lens mass and redshift are $\Mvir=10^{14}h^{-1}\Msun$ and $z_l=0.2$, the 
source is at redshift $z_s=1.0$ and the halo 
is assumed to have a NFW profile. 
({\it Right}): The virial radius $\thetavir$ in angular units as a 
function of $z_l$ for different halo masses. 
}
\label{fig:kappa}
\end{figure*}

\begin{figure*}
\resizebox{80mm}{!}{\includegraphics[angle=0]{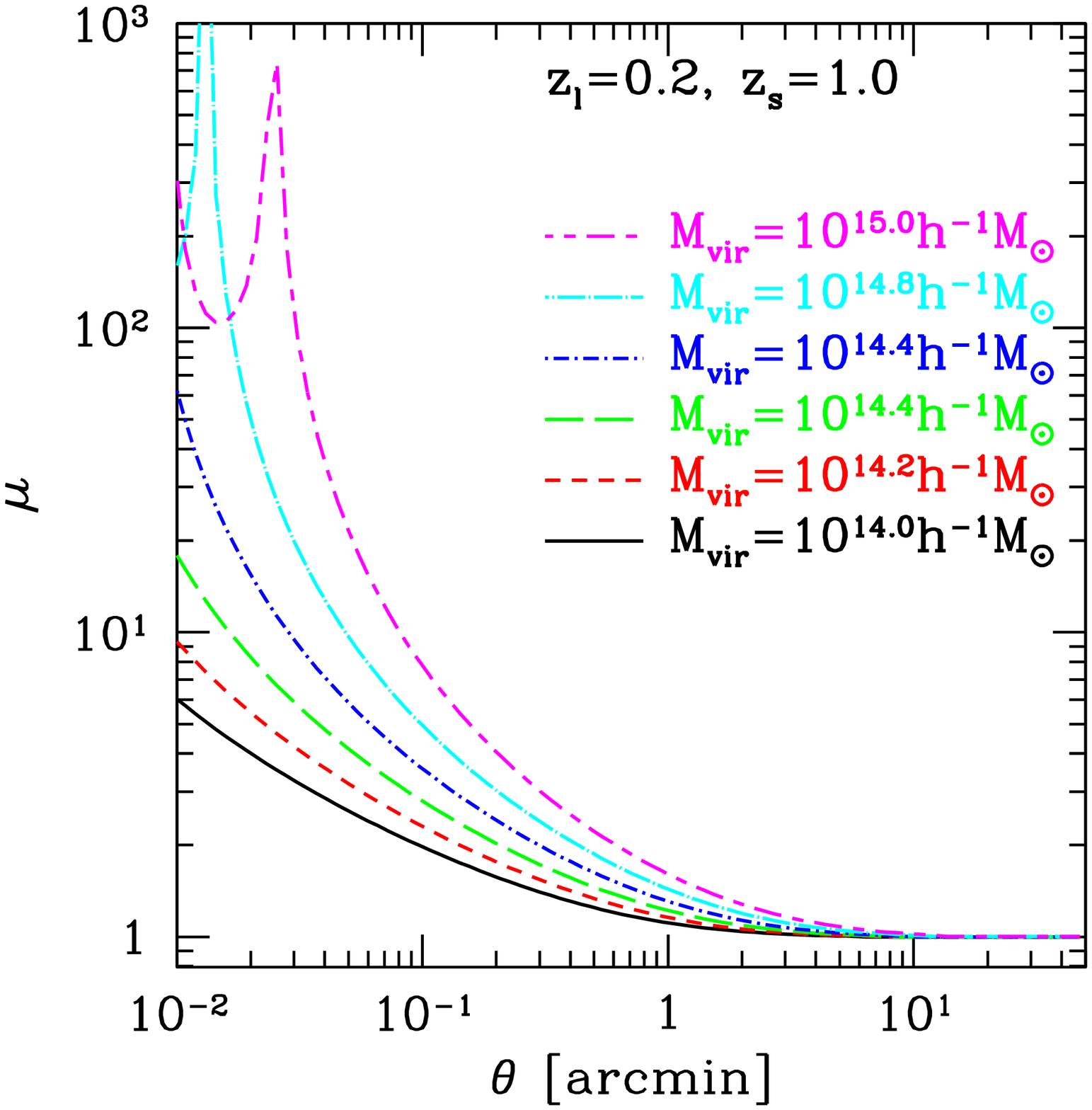}}
\resizebox{80mm}{!}{\includegraphics[angle=0]{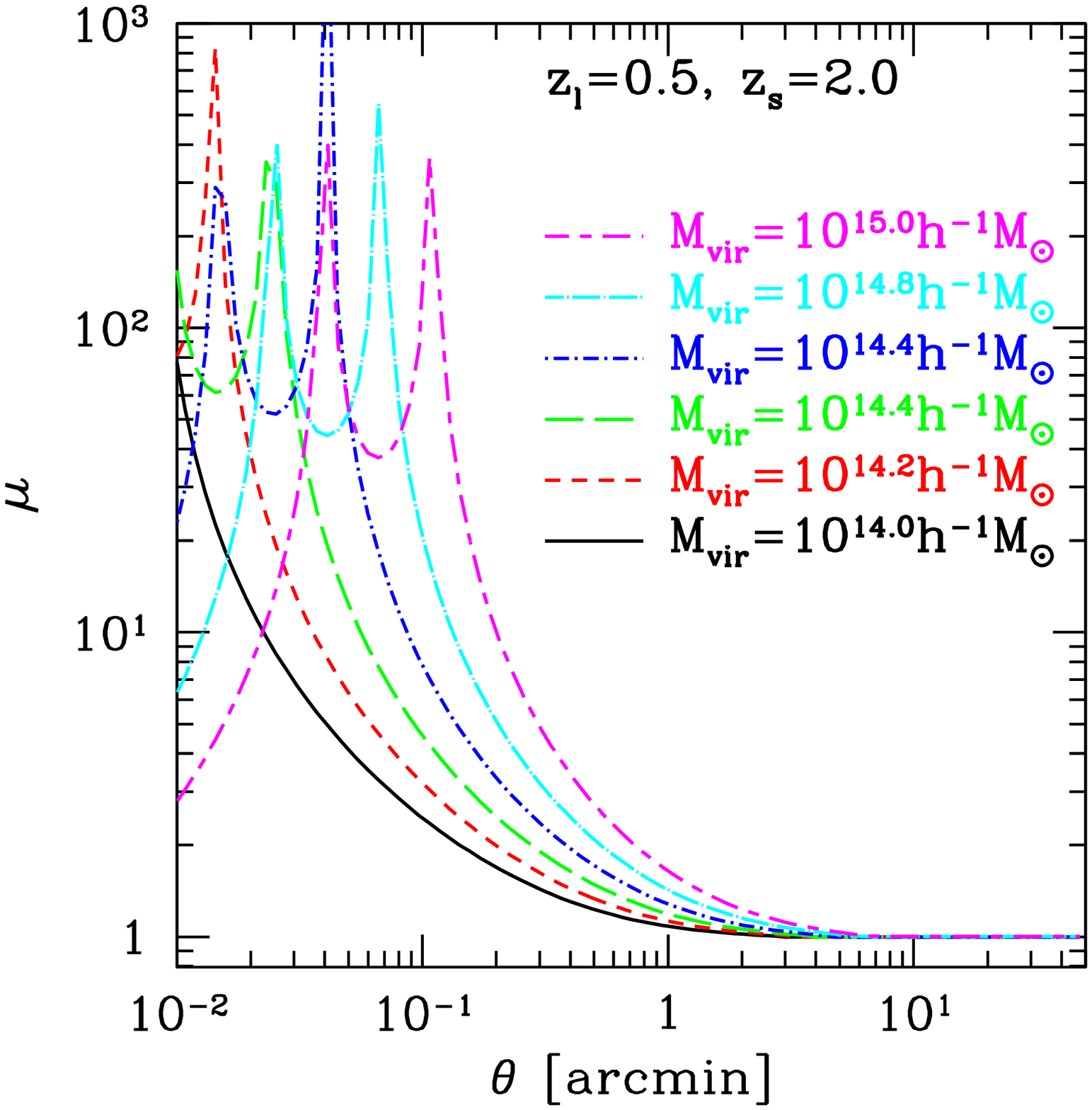}}
  \resizebox{80mm}{!}{\includegraphics[angle=0]{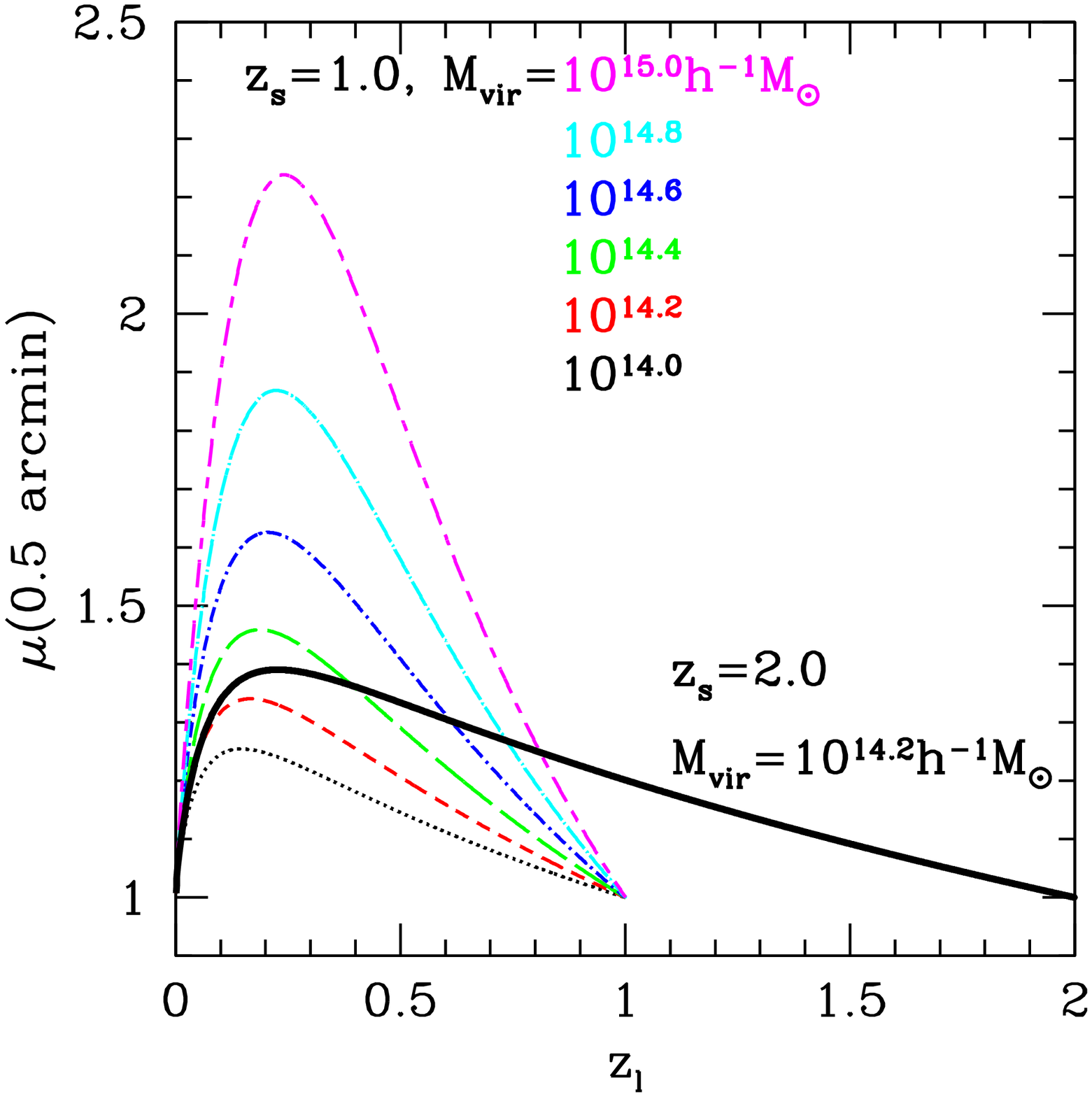}}
  \resizebox{80mm}{!}{\includegraphics[angle=0]{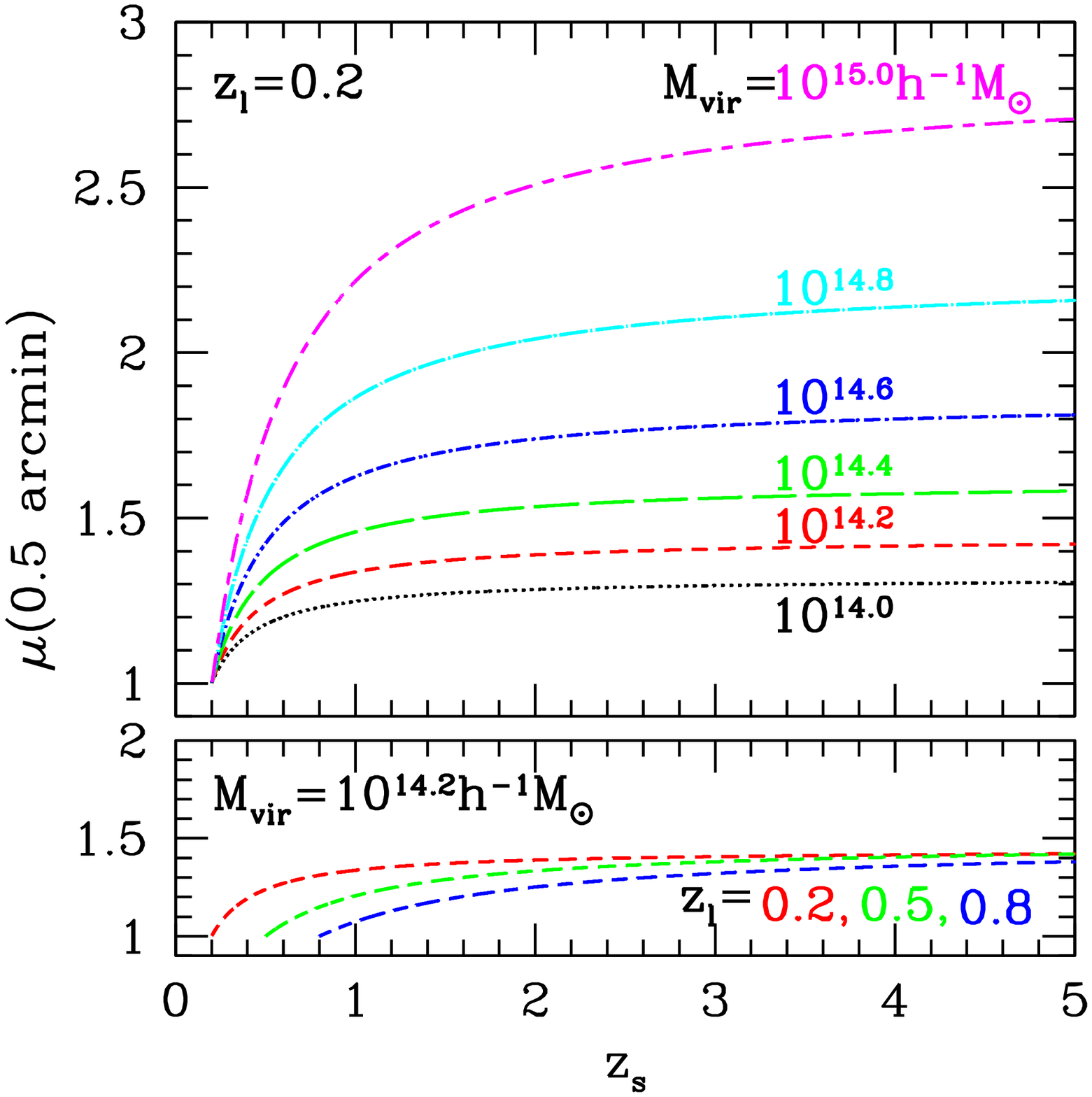}}
  \caption{({\it Top Left}): Magnification $\mu$ 
as a function of angular separation $\theta$ from the cluster center, assuming
a NFW profile, for different cluster masses with $z_l=0.2$ and $z_s=1.0$.
({\it Top Right}): Same but for $z_l=0.5$ and $z_s=2.0$. 
({\it Bottom Left}): $\mu$ evaluated at $\theta=0.5$~arcmin as a function of lens 
redshift $z_l$ for $z_s=1.0$ at different cluster masses and for $z_s=2.0$ 
for cluster mass $\Mvir=10^{14.2}h^{-1}M_{\odot}$.
({\it Bottom Right}): $\mu(\theta=0.5\rm{arcmin})$ as a function of 
source redshift $z_s$ for $z_l=0.2$ at different cluster masses (upper panel) and for 
$\Mvir=10^{14.2}h^{-1}M_{\odot}$ at different values of $z_l$ (lower panel).}
\label{fig:mu}
\end{figure*}

Given the 3d density profile $\rho(r)$, we can compute all lensing quantities of interest
in the lens plane by performing successive numerical integrations and differentiations. 
For the NFW profile, most quantities can be computed analytically, speeding up
calculations that use the lensing properties. The projected density field $\Sigma$ is obtained 
integrating over the parallel coordinate $\rpara=\chi$ in the
position vector decomposed as ${\bf x}=(\rpara,\rperp)$, where the perpendicular coordinate 
$\rperp=D_A(\chi)\theta$, $D_A$ is the angular diameter distance and $\theta$ is the angular
coordinate in the lens plane. 
Here and throughout, all distances are {\it comoving}.
For the NFW profile the projected surface density is given by \citep{TakJai03a}
\begin{eqnarray}
\Sigma(\theta)=\int_{-\rvir}^{\rvir}d\rpara \rho(r) = \frac{\Mvir fc^2}{2\pi\rvir^2}F(c\theta/\thetavir)\,,
\end{eqnarray}
where
\begin{eqnarray}
F(x)=
\left\{
\begin{array}{ll}
-\frac{\sqrt{c^2-x^2}}{(1-x^2)(1+c)}
+\frac{1}{(1-x^2)^{3/2}}{\rm arccosh}\frac{x^2+c}{x(1+c)}\,,
&(x<1)\\
\frac{\sqrt{c^2-1}}{3(1+c)}\left[1+\frac{1}{c+1}\right]\,,
&(x=1)\\
-\frac{\sqrt{c^2-x^2}}{(1-x^2)(1+c)}
-\frac{1}{(x^2-1)^{3/2}}{\rm arccos }\frac{x^2+c}{x(1+c)}\,,
&(1<x<c)\\
0\,,
&(x>c)
\end{array}
\right.
\end{eqnarray}
and $\thetavir(\chi)=\rvir/D_A(\chi)$.
The convergence field $\kappa$ is defined in terms of the critical
density $\Sigma_{\rm crit}$ by
\begin{eqnarray}
\kappa(\theta)&=&\frac{\Sigma(\theta)}{\Sigma_{\rm crit}}, \ \ \ 
{\Sigma_{\rm crit}} = \frac{a}{4\pi G}                        
                       \frac{D_A(\chi_s)}{D_A(\chi) D_A(\chi_s - \chi)}\,.
\end{eqnarray}
where $a=(1+z)^{-1}$ is the scale factor.
The convergence measures isotropic light distortions and is related
to the projected lensing potential $\varphi$ via the Poisson equation 
$\nabla^2\varphi=2\kappa$. 
Anisotropic distortions
are measured by the complex shear field $\vec{\gamma}=\gamma_1+i\gamma_2$, whose amplitude 
for an {\it axially symmetric} mass distribution is simply related
to the convergence by
\begin{eqnarray}
\gamma(\theta)&=&\bar{\kappa}(<\theta)-\kappa(\theta)\,,
\end{eqnarray}
where the average convergence field up to $\theta$ is
\begin{eqnarray}
\bar{\kappa}(<\theta)&=&\frac{1}{\pi \theta^2}\int_{0}^{\theta} 
                        d\theta^{\prime} 2\pi\theta^{\prime} \kappa(\theta)\,.
\end{eqnarray}

Like the convergence, the shear can be computed analytically for 
the NFW profile and is given by \citep{TakJai03b}
\begin{eqnarray}
\gamma(\theta)= \frac{\Mvir fc^2}{2\pi\rvir^2}\frac{G(c\theta/\thetavir)}{\Sigma_{\rm crit}} \,,
\end{eqnarray}
where
\begin{eqnarray}
G(x)=
\left\{
\begin{array}{ll}
\frac{1}{x^2(1+c)}\left[\frac{(2-x^2)\sqrt{c^2-x^2}}{1-x^2}-2c\right]
+\frac{2}{x^2}\ln \frac{x(1+c)}{c+\sqrt{c^2-x^2}}
+\frac{2-3x^2}{x^2(1-x^2)^{3/2}}{\rm arccosh}\frac{x^2+c}{x(1+c)}\,,
&(x<1)\\
\frac{1}{3(1+c)}\left[\frac{(11c+10)\sqrt{c^2-1}}{1+c}-6c\right]
+2\ln \frac{1+c}{c+\sqrt{c^2-1}}\,,
&(x=1)\\
\frac{1}{x^2(1+c)}\left[\frac{(2-x^2)\sqrt{c^2-x^2}}{(1-x^2)}-2c\right]
+\frac{2}{x^2}\ln \frac{x(1+c)}{c+\sqrt{c^2-x^2}}
-\frac{2-3x^2}{x^2(x^2-1)^{3/2}}{\rm arccos }\frac{x^2+c}{x(1+c)}\,,
&(1<x<c)\\
\frac{2f^{-1}}{x^2}\,.
&(x>c)
\end{array}
\right.
\end{eqnarray}

Finally, the magnification $\mu$ measures the total flux amplification as well as the 
increase in angular size of source galaxies, and is given 
by the inverse determinant of the Jacobian transformation between source and image 
angular coordinates. It is expressed in terms of the convergence and shear as
\begin{eqnarray}
\mu(\theta)=\frac{1}{\left[1-\kappa(\theta)\right]^2-|\gamma(\theta)|^2}\,. 
\label{eq:mag}
\end{eqnarray}

In Fig.~\ref{fig:kappa}, we show various lensing observables, assuming
a spherical  
NFW profile lens halo of mass $\Mvir=10^{14}h^{-1}M_{\odot}$ at redshift
$z_l=0.2$ and a source galaxy at  
$z_s=1.0$. The convergence tends to dominate over the shear in the halo core 
whereas the shear dominates in the outer parts. 
Likewise the magnification is close to 1 (no magnification) far from the halo 
and rises to significant values close to the cluster center. 
Also shown is the virial radius in angular units $\thetavir$ as a function 
of halo redshift for different halo masses. 
Halos of interest have typical angular sizes of the order of few arcminutes.  

In Fig.~\ref{fig:mu}, we show the magnification $\mu$ 
as a function of angular separation $\theta$, lens redshift $z_l$ and source
redshift $z_s$ for different cluster masses. 
As the halo mass increases the magnification is significantly enhanced. 
The two spikes in each magnification
curve at the top panels of Fig.~\ref{fig:mu} represent critical curves 
(tangential and radial) where the magnification 
is formally infinite and correspond to the two solutions of the quadratic 
equation $|1-\kappa|^2=\gamma^2$.

Magnification increases with lens redshift, reaching a maximum, and then 
decreases as the lens approaches the source; 
that reflects the lensing efficiency of the critical 
surface density terms. 
Magnification also increases with source redshift, until it reaches a
plateau; therefore one expects roughly similar lensing magnification effects 
from sources at sufficiently high redshifts.

In what follows, we use the letters $\kappa$, $\gamma$ and $\mu$ to 
denote {\it full} lensing quantities, which incorporate halo ellipticity (\S~\ref{sub:ellip}). 
The spherically symmetric results of this section will be denoted by the same letters 
with a tilde (e.g. $\tilde{\kappa}$).

\subsection{Elliptical Lenses} \label{sub:ellip}

Ellipticity can be introduced in a number of ways into halo profiles. The most obvious
choice is to introduce it directly into the 3d density profile $\rho(r)$ and project it
to obtain $\Sigma$. 
Another choice is to introduce ellipticity in $\Sigma$ directly. 
The computation of the lensing quantities can then be obtained numerically by performing
integrations (of $\Sigma$) to obtain deflection angles, and 
differentiations (of deflection angles) to obtain shear and 
magnification fields. 
In special cases of known potential-density pairs, these quantities can be computed 
analytically. 

Another possibility is to introduce ellipticity in the two-dimensional 
projected lensing potential $\varphi$, 
whose second derivatives give the lensing quantities more directly. 
The latter approach may produce density profiles with dumbbell 
shapes for high values of ellipticity, but produces physical 
magnification fields for low/intermediate ellipticities close to those
expected for the majority of halos \citep{JinSut02}. 
Moreover, this approach leads to simple analytical expressions that can 
be expressed in terms of the usual results of the spherically symmetrical 
case and, for this reason, this is the choice we implement.

\begin{figure}
\resizebox{180mm}{!}{\includegraphics[angle=-90]{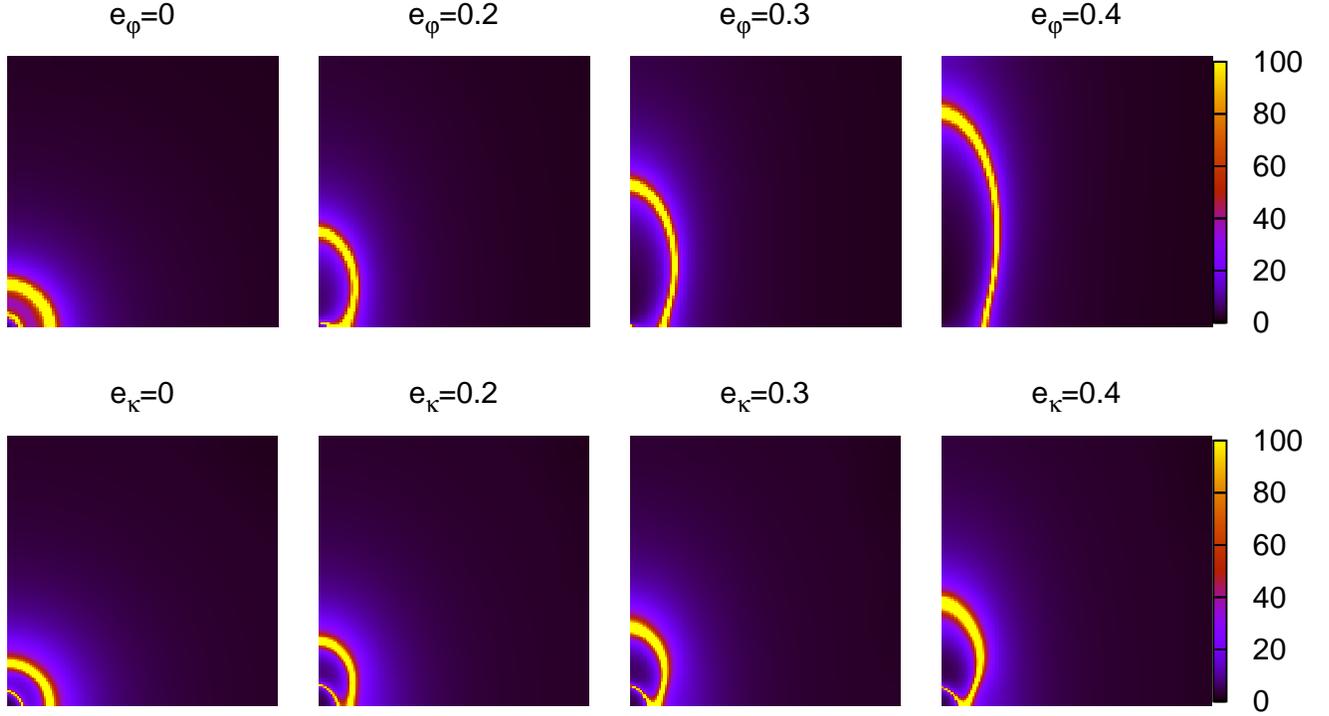}}
  \caption{ Maps of magnification $\mu$ for different values of ellipticity 
with $z_l=0.5$ and $z_s=2.0$ and halo mass 
$\Mvir=10^{15}$~$h^{-1}M_{\odot}$. Here $e=1-b/a$, where $a$ and $b$ are the
projected major and minor ellipse axes in the lens plane. 
({\it Top panels}): Ellipticity $e_{\varphi}$ is introduced 
directly in the {\it potential} profile  
using the analytical expressions derived in this paper.
({\it Bottom panels}):
Ellipticity $e_{\kappa}$ is introduced in the {\it density} profile 
and propagated to the deflection angle and magnification numerically 
using the WSLAP code of Diego et al (2007). 
Notice that $e_{\varphi}=0.2$ produces roughly the same magnification map
as $e_{\kappa}=0.4$. The field of view is 0.6~arcmin in size. 
}
\label{fig:map_mu}
\end{figure}

\subsubsection{Elliptical Potential}

\begin{figure*}
\resizebox{88mm}{!}{\includegraphics[angle=0]{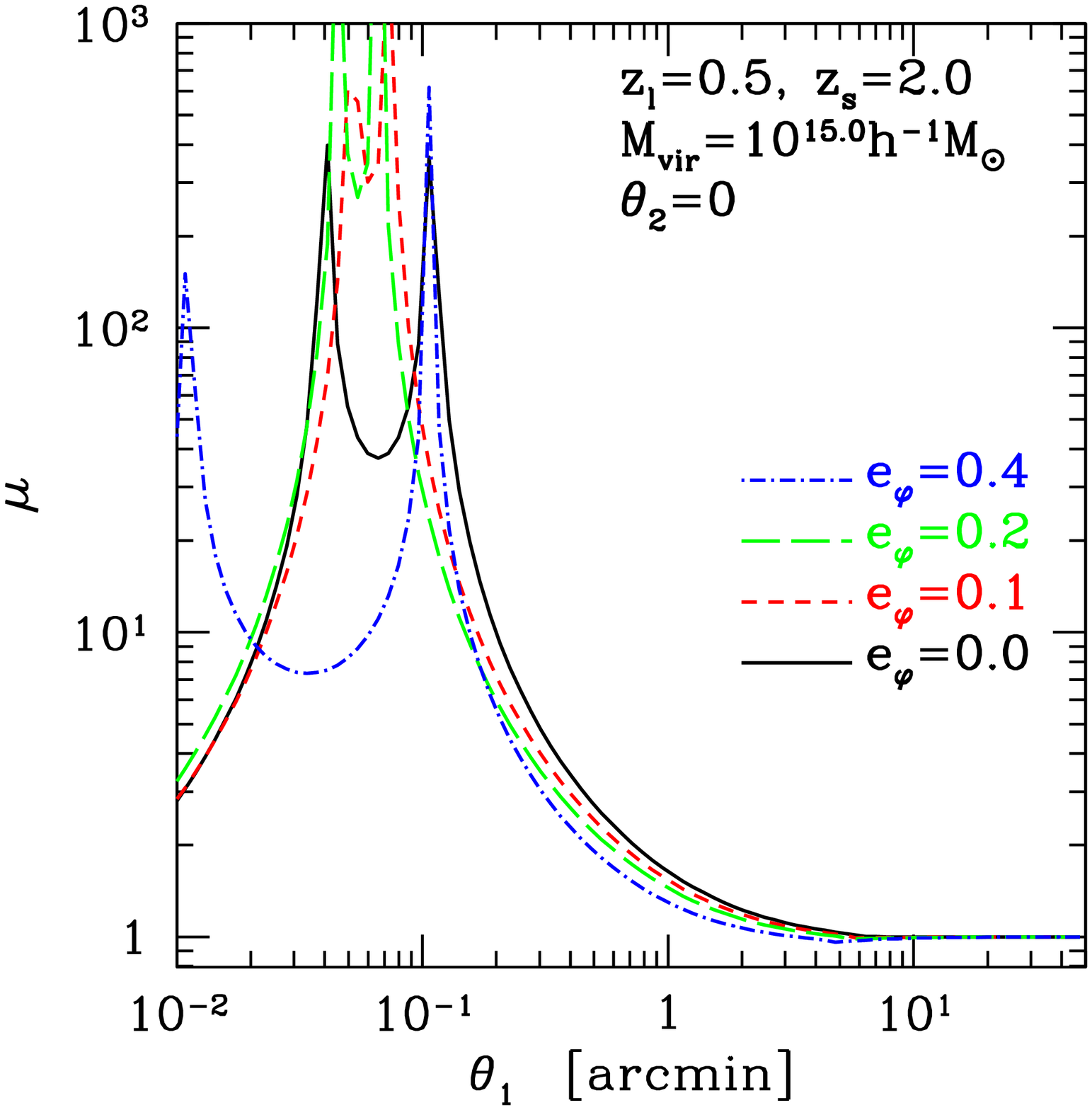}}
\resizebox{88mm}{!}{\includegraphics[angle=0]{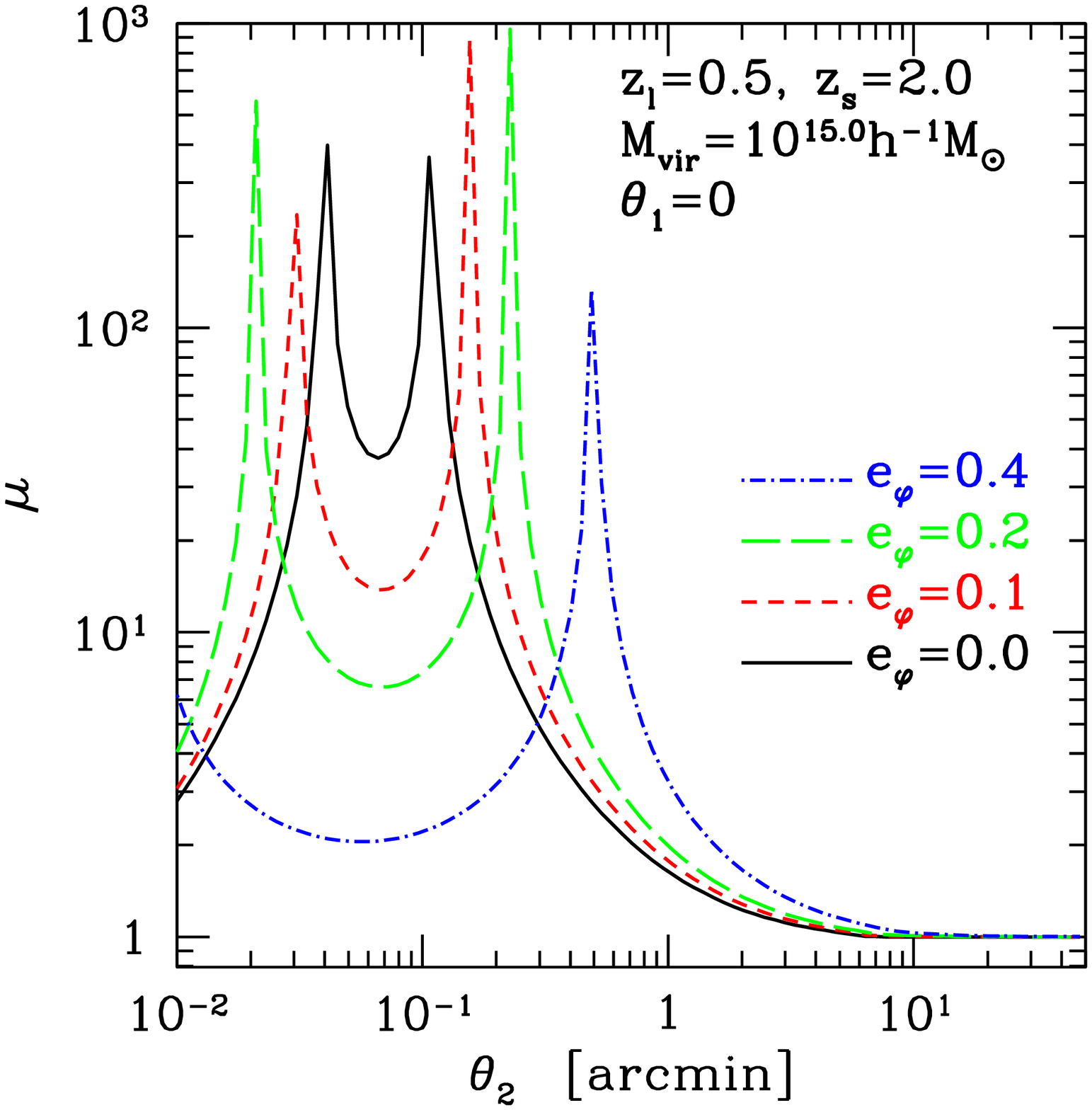}}
  \caption{ ({\it Left}): Magnification $\mu$ 
as a function of angular separation from the cluster center in the 
$\theta_1$-direction ($\theta_2=0$). 
({\it Right}): Same but in the 
$\theta_2$-direction ($\theta_1=0$). We assume
an elliptical NFW profile, with different values of ellipticity, 
$z_l=0.5$ and $z_s=2.0$. Here $\ephi=1-b/a$, where $a$ and $b$ are the
major and minor ellipse axes and ellipticity is introduced in the lensing
potential in the
$\theta_2$-direction with the prescription of Meneghetti et al. (2003).}
\label{fig:mu_ellip}
\end{figure*}

Given an axially symmetric lensing potential $\varphi(\theta)$, we follow 
\cite{MenBarMos03} and obtain
the elliptical generalization with major axis along the $\theta_2$ direction 
by substituting $\theta^2=\theta_1^2+\theta_2^2$ by
\begin{eqnarray}
\theta\rightarrow\bar{\theta}=\sqrt{\frac{\theta_1^2}{(1-e_{\varphi})}+\theta_2^2(1-e_{\varphi})}\,,
\end{eqnarray}
where $e_\varphi=1-b/a$ with $a$ and $b$ being the major and minor ellipse axes. 
Our approach is similar to that of \cite{GoeKne02}, though our NFW profiles
are truncated at the virial radius as opposed to extending to infinity.
We can then compute the deflection angle components, their derivatives 
and the lensing quantities of interest for elliptical lenses. 
In Appendix A, we express the results for $\kappa$, $\gamma_1$ and
$\gamma_2$ in terms of the spherically symmetric case, 
with the effects of ellipticity described analytically.
Using the expressions given in Eqs.~\ref{eq:kappa_elipphi}, 
\ref{eq:gamma1_elipphi} and \ref{eq:gamma2_elipphi} we can compute the 
magnification defined by Eq.~(\ref{eq:mag}). 

In Fig.~\ref{fig:map_mu}, we show
magnification maps in the $\theta_1 \times \theta_2$ plane, for
different ellipticities. The top row shows maps derived using 
our model in which ellipticity is introduced in the lensing
potential. The bottom row shows maps derived using the WSLAP code 
\citep{DieTegProSan07}, where ellipticity $e_\kappa$ is introduced in
the 3d density profile,  
which is then projected. Notice that both agree when
$e_\varphi=e_\kappa=0$ and that the ellipticity  
introduced in the potential causes larger changes in the magnification map.
This is expected, since the second derivative of the potential is
related to the density -- the quantitative connection between the two is
presented below.

In Fig.~\ref{fig:mu_ellip}, we show the magnification $\mu$ 
as a function of angular separation along the two directions $\theta_1$ and $\theta_2$ 
for different values of $e_\varphi$. In the $\theta_2$ direction of the major axis, the
tangential caustic moves to larger angles whereas the radial caustic
moves to smaller angles. 
In the $\theta_1$ direction much less dramatic changes happen, as also seen in 
Fig.~\ref{fig:map_mu}.

\subsubsection{Elliptical Density versus Elliptical Potential}

\begin{figure}
\resizebox{180mm}{!}{\includegraphics[angle=-90]{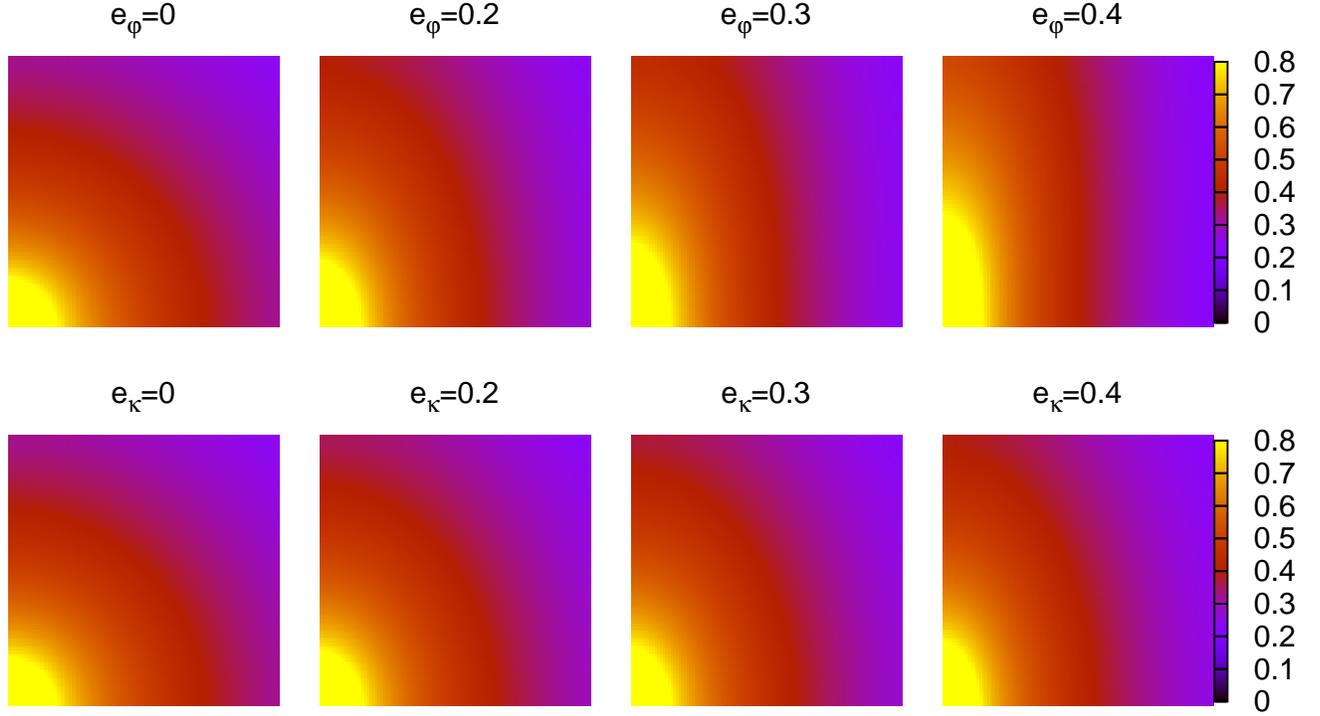}}
  \caption{ Maps of convergence $\kappa$ for different values of 
ellipticity with halo redshift $z_l=0.5$ and $z_s=2.0$ and halo mass 
$\Mvir=10^{15}$~$h^{-1}M_{\odot}$. 
Ellipticities are again introduced in the 
{\it potential} profile ({\it Top panels}) and in the 
{\it density} profile ({\it Bottom panels}). 
The field of view is 0.6~arcmin in size.
}
\label{fig:map_kappa}
\end{figure}

Simulations typically predict ellipticities in the halo density profile. Therefore it is 
interesting to relate the ellipticity introduced in the potential as in the previous
section, denoted $e_{\varphi}$, to the corresponding effective ellipticity in the 
surface density profile, denoted $e_{\kappa}$, introduced directly in the 
convergence map instead of the potential, i.e. 
by substituting $\theta^2=\theta_1^2+\theta_2^2$ in $\kappa$ by
\begin{eqnarray}
\theta\rightarrow \theta^{*}=\sqrt{\frac{\theta_1^2}{(1-\ek)}+\theta_2^2(1-\ek)}\,.
\end{eqnarray}

\begin{figure*}
\resizebox{88mm}{!}{\includegraphics[angle=0]{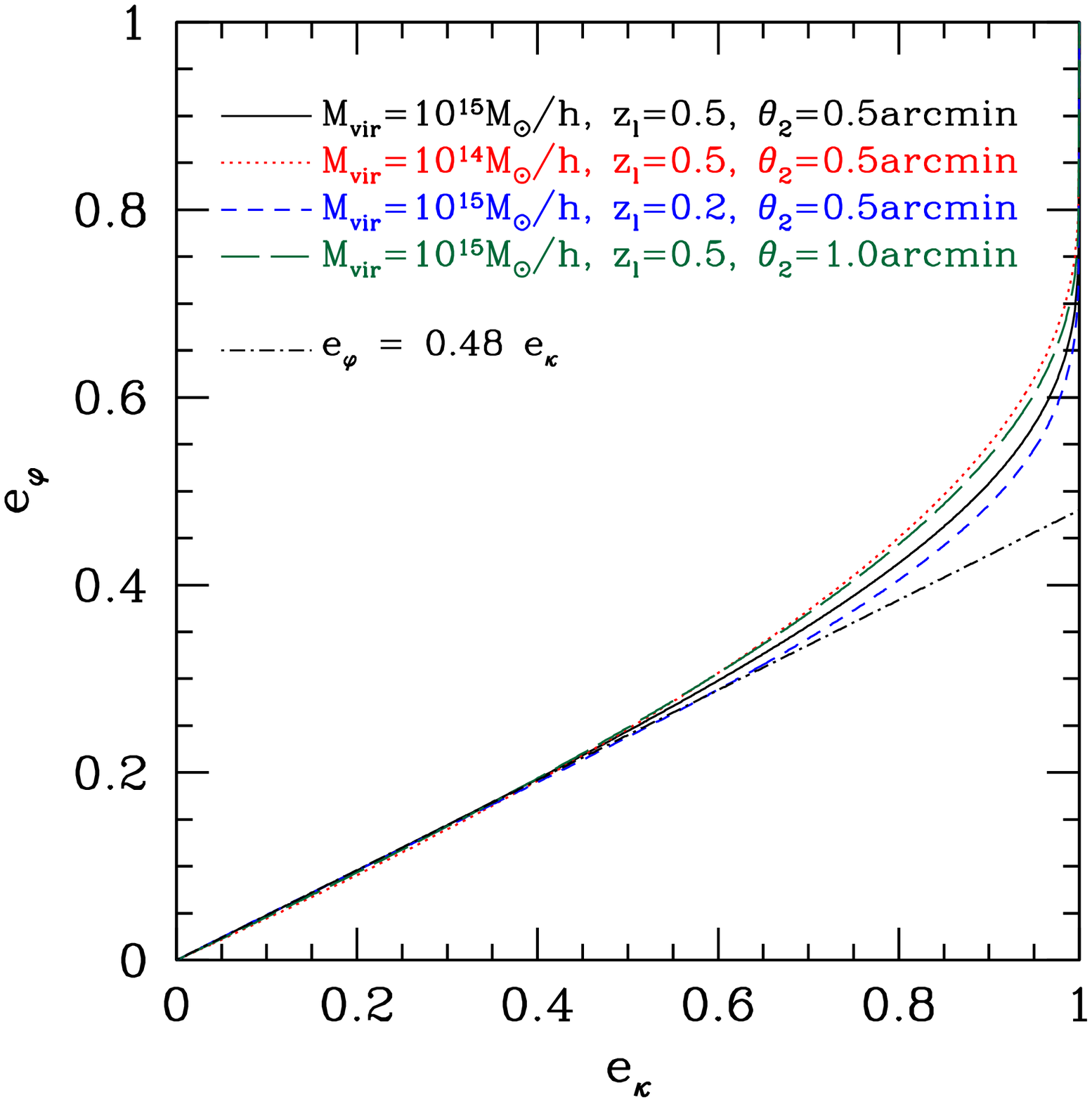}}
  \caption{ Relationship between $\ephi$ and $\ekappa$ which produces the 
same value of the convergence at the matching point $(\theta_1=0,\theta_2)$,
shown for 
different values of halo mass, redshift and matching coordinate $\theta_2$. 
Because the potential field is smoother than the density field, values of
$\ephi$ which are only about half those of $\ekappa$ produce 
the same convergence field for $\ekappa<0.5$. 
}
\label{fig:ephi_ekappa}
\end{figure*}

In Fig.~\ref{fig:map_kappa} we show convergence maps for different 
ellipticities, similarly to Fig.~\ref{fig:map_mu}. Again,
ellipticity $\ephi$ is introduced in the lensing potential (top row) 
and $\ekappa$ in the density (bottom row). 
Fig.~\ref{fig:ephi_ekappa} shows one way to relate the two kinds of
ellipticity. Enforcing
the match $\kappa(\theta_1,\theta_2)=\tilde{\kappa}(\theta^{*})$, 
it shows the relationship between $\ek$ and $\ephi$. 
Here $\kappa(\theta_1,\theta_2)$ is the convergence that results
from adding $\ephi$ in the lensing potential and $\tilde{\kappa}$
is the spherically symmetric convergence. 
Specifically we match the value of the convergences 
at the arbitrary matching point
$(\theta_1=0,\theta_2)$ for various choices of cluster mass, 
lens and source redshift and matching coordinate $\theta_2$.
The relation is quite insensitive to these parameters and also
to the value of $\theta_2$ chosen for the match.
Overall the relation $\ephi=0.48\ekappa$ seems to hold
relatively well for $\ekappa<0.5$ in the range of parameters considered here.
This is consistent with the results of \cite{GoeKne02} for the
extended NFW profile, where they find $\ekappa \sim 2\ephi$ for
$\ephi < 0.25$.
Our results can be used to relate
the typical values of $\ek$ from simulations to the more convenient values of 
$\ephi$ used to analytically model the effects of ellipticity.

\section{Lensing Statistics} \label{sec:prob}

\subsection{Halo Counts}

Given the differential comoving number density of lens halos 
$dn/d\ln \Mvir$, one can estimate the cluster number counts 
$d N$ in the redshift/mass range $d z_l d \ln \Mvir$ and 
solid angle $\Delta \Omega$ in the sky as
\begin{eqnarray}
d N(z_l,\Mvir)=d z_l d\ln \Mvir \frac{dV(z_l)}{dz_l} \frac{dn(z_l,\Mvir)}{d\ln \Mvir}\,, \\ \nonumber
\end{eqnarray}
where $dV/dz_l=\Delta \Omega D_A^2/H$ is the comoving volume element, 
$D_A(z_l)$ is the comoving angular
diameter distance and $H(z_l)$ is the Hubble parameter.
The differential halo density $n$ and counts $N$ above a mass 
threshold $M_{\rm th}$ are given by
\begin{eqnarray}
n(z_l) &=& 
      \int_{M_{\rm th}}^{\infty} d\ln \Mvir   
      \frac{dn(z_l,\Mvir)}{d\ln \Mvir}\,, \\ \nonumber
\frac{dN(z_l)}{dz_l} &=& 
      \Delta\Omega \frac{D_A^2(z_l)}{H(z_l)}n(z_l)\,.
\end{eqnarray}

\subsection{Lensing Cross-Section}

Each halo produces a elliptical ``ring'' region around it, with area 
$\Delta \Omega_{\mu}(z_l,z_s,M,\ephi,\mu_{min})$, 
such that inside this area the magnification is larger 
than $\mu_{min}$. This area is an effective cross-section for lensing
statistics. 
Note $\Delta \Omega_{\mu}$ is the area with 
$\mu>\mu_{min}$ in the source plane
\begin{eqnarray}
\Delta \Omega_{\mu}(\mu_{min}) = \int_{\mu >\mu_{min}} d\beta^2 
                               = \int_{\mu(\theta)>\mu_{min}} \frac{d\theta^2}{\mu(\theta)}\,,
\end{eqnarray}
where we used the fact that the magnification is precisely the Jacobian of the
transformation between image and source coordinates $\mu=d\theta^2/d\beta^2$. 
A reference value for $\Delta \Omega_{\mu}$ is the projected cluster surface
area defined by its virial radius in the image plane 
$\Delta \Omega_{\rm vir}=\pi \thetavir^2$. 
For large values of $\mu$ we expect the magnification region to be well inside the halo
core and therefore $\Delta  \Omega_{\mu} \ll \Delta \Omega_{\rm vir}$, whereas for 
$\mu \rightarrow 1$, $\Delta \Omega_{\mu}$ becomes formally infinite.  
For a given halo, we estimate $\Delta \Omega_{\mu}$ by 
evaluating $\mu$ in a square grid of $100 \times 100$ points with side $2\thetavir$.
We then multiply the fraction of grid points with $\mu>\mu_{min}$ by 
the area $4\thetavir^2$; in this fraction, each point is weighted by  
$1/\mu(\theta)$ at the grid. 

Note that the way we estimate $\Delta \Omega_\mu$ is only strictly accurate 
if $\Delta \Omega_\mu < 4\thetavir^2$. This is not true for sufficiently 
low $\mu$, where $\Delta \Omega_\mu$ formally covers the whole sky.
In principle this estimation can be improved by adaptively surveying a 
radius larger than $2\thetavir$ around the halo. For the goal of estimating
the lensing probability, this is not necessary as discussed in the next 
two sections.

\begin{figure*}
\resizebox{58mm}{!}{\includegraphics[angle=0]{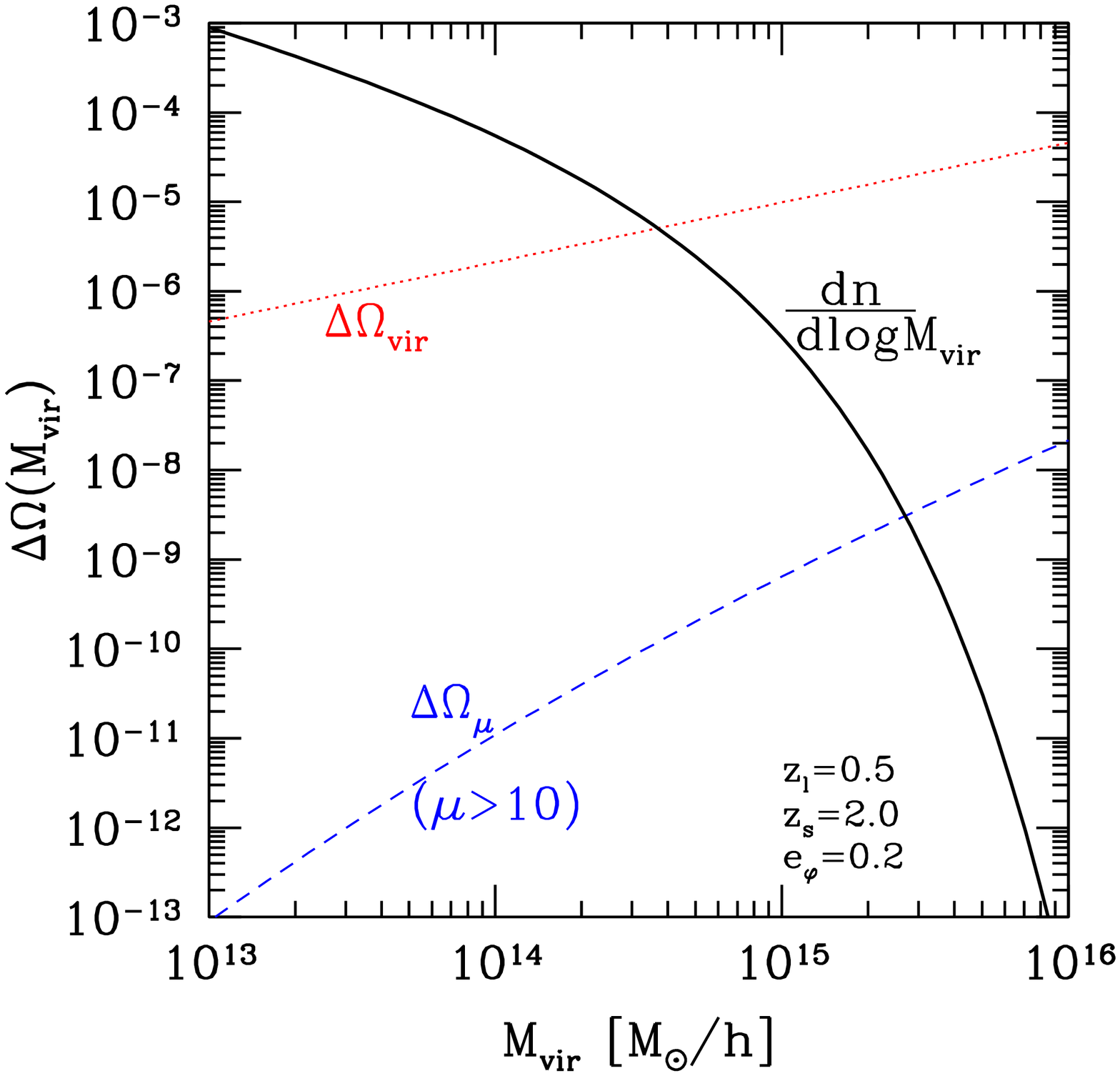}}
\resizebox{58mm}{!}{\includegraphics[angle=0]{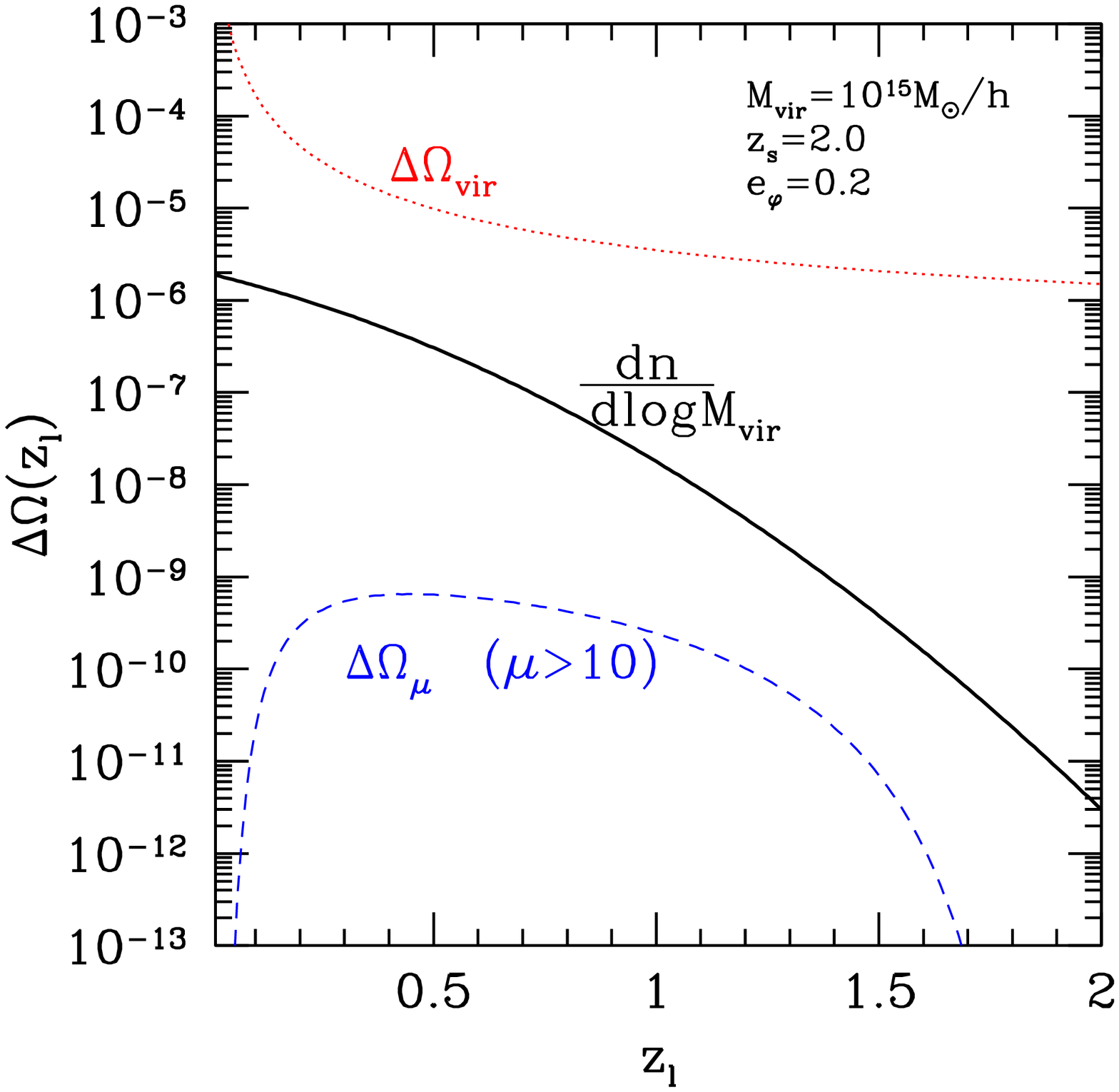}}
\resizebox{58mm}{!}{\includegraphics[angle=0]{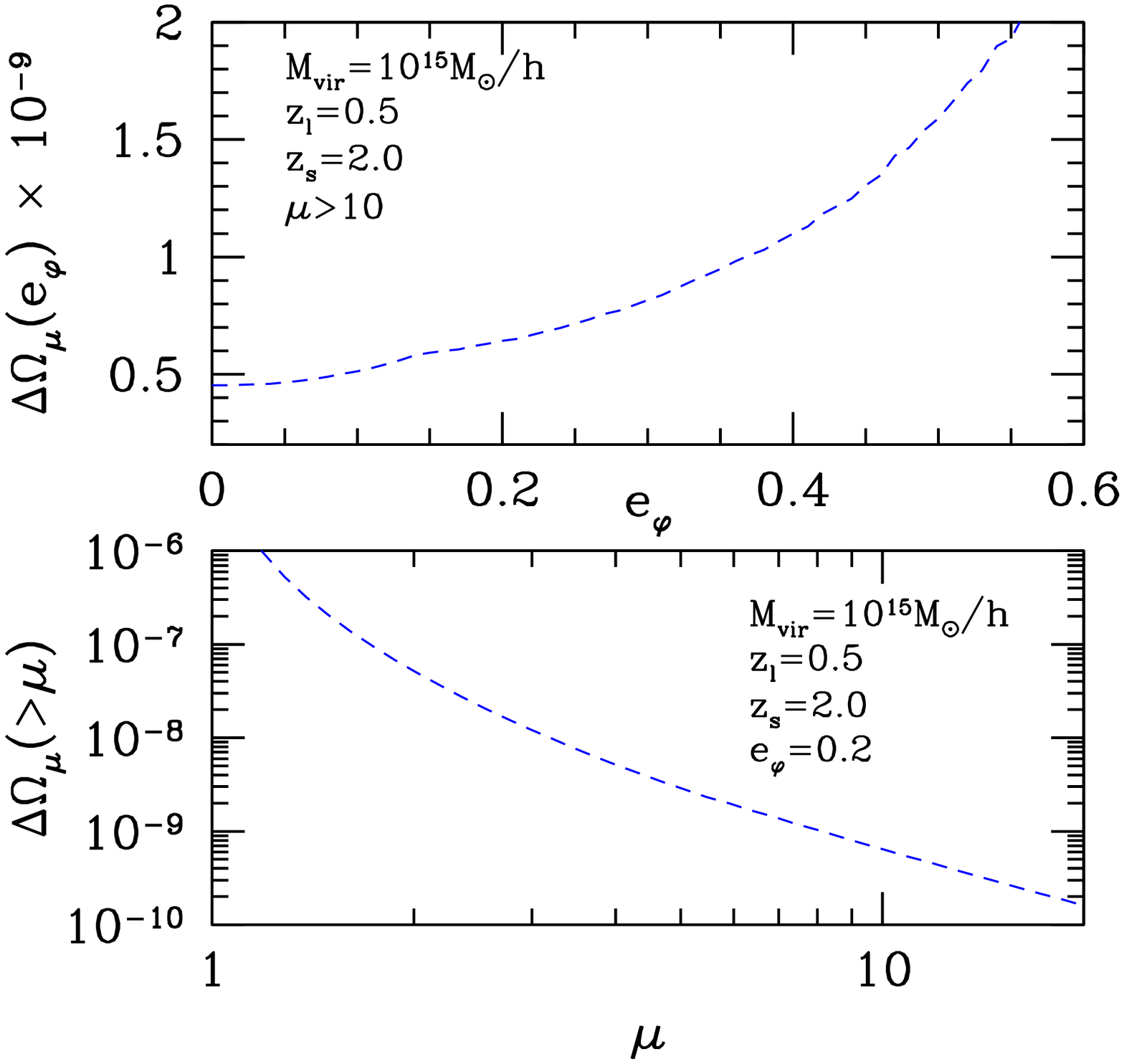}}
  \caption{ Virial and lensing cross-sections (in rad$^2$) 
as a function of halo mass, redshift, ellipticity and minimum magnification. 
({\it Left}): Both $\Delta\Omega_{\mu}$ and $\Delta\Omega_{\rm vir}=\pi \thetavir^2$ as a function of 
halo mass $\Mvir$ for $z_l=0.5$, $z_s=2.0$, $\ephi=0.2$ and $\mu_{min}=10$. 
({\it Middle}): Same but as a function of $z_l$ for $\Mvir=10^{15}h^{-1}M_{\odot}$.
({\it Right}): $\Delta\Omega_{\mu}$ versus $\ephi$ for $\mu_{min}=10$ (upper panel) and versus $\mu$
for $\ephi=0.2$ (lower panel).
 }
\label{fig:Omegas}
\end{figure*}

In Fig.~{\ref{fig:Omegas}} we show $\Delta\Omega_{\mu}$ and $\Delta\Omega_{\rm vir}$
as a function of halo mass, redshift, ellipticity and minimum magnification. 
Since $\rvir \sim \Mvir^{1/3}$, we have that $\Delta\Omega_{\rm vir} \sim \Mvir^{2/3}$. 
For this high value of $\mu_{\rm min}$, $\Delta\Omega_{\mu}$ is many orders of 
magnitude smaller but growing faster with $\Mvir$. 
Since $\rvir \sim [\Delta_c(z)E^2(z)]^{-1/3}$, we have that 
$\Delta\Omega_{\rm vir} \sim \Delta_c(z)^{-2/3}E(z)^{-4/3}D_A(z)^{-2}$. 
The redshift dependence of $\Delta\Omega_{\rm vir}$ is dominated by
that of $D_A$, since $\Delta_c(z)$ and $E(z)$ change less. 
In particular, it increases rapidly at
low redshifts, where $D_A \rightarrow 0$.
On the other hand $\Delta\Omega_{\mu}$, which
shows a similar trend at intermediate redshifts, 
goes to zero as $z_l \rightarrow 0$ or $z_s$, where the lensing efficiency 
vanishes. 

As also apparent from the magnification maps, there is a relatively weak 
dependence on $\ephi$, where ellipticity mainly distorts the critical curves, but do 
not change their area.
A strong dependence on $\mu_{min}$ can be seen; the largest areas come from 
$\mu_{min}\simlt 2$ even for the most massive halos.  We will see below that
once we sum over halo mass, the contribution of high magnification 
($\mu\simgt 2$) regions to observable number counts is very small. 

\subsection{Sky fraction with large magnifications: Lensing optical depth}

We can estimate the 
fraction $f_{\mu}$ of the sky with $\mu>\mu_{min}$ due to all halos 
above a certain mass and redshift range as 
\begin{eqnarray}
f_{\mu}&=& \int_{0}^{z_s} dz_l \frac{D_A^2(z_l)}{H(z_l)} 
          \int_{M_{\rm th}}^{\infty} d\ln \Mvir 
          \int_{0}^{\infty} dz_s P(z_s) 
          \int_{0}^{1} d\ephi P(\ephi)
          \Delta \Omega_{\mu}(z_l,z_s,\Mvir,\ephi,\mu_{min}) 
          \frac{dn(z_l,\Mvir)}{d\ln \Mvir}\,,
\end{eqnarray}
where $P(z_s)$ is the source redshift distribution and $P(\ephi)$ is the 
distribution of halo ellipticities. 
This quantity is the optical depth for lensing magnification and 
determines the probability that a given source galaxy/population is 
magnified by intervening lens halos. This definition of $f_\mu$ accounts 
for overlapping angular regions, which are counted multiple times in 
the integrals in mass and redshift.
However it neglects multiple lens events, such as multiple images, since 
only one event is counted per halo. 
Therefore, strictly speaking, $f_\mu$ is the sky area with large 
magnifications only for $f_\mu \ll 1$, i.e. $\mu \gg 1$, where no such 
overlappings occur.

Note also that $f_\mu$ can be, and in fact is, larger than 1 at regions 
with $\mu \sim 1$, i.e. such regions cover the sky many times.
In fact, even $\Delta \Omega_\mu$ itself for a given cluster can be as 
large as the whole sky for $\mu \sim 1$.
How large $f_\mu$ becomes as $\mu$ approaches 1 depends on the top value 
we allow $\Delta \Omega_\mu$ to have, which in our case was set as 
$4\thetavir^2$ and causes a saturation in $f_\mu$ at $\sim 27$ for $\mu \le 1$. 
Even though this top cut-off in $\Delta \Omega_\mu$ affects our computation 
of $f_\mu$ for very low values of $\mu$, it does not significantly change the 
quantity we are really interested in, the lensing probability $P(>\mu)$ 
described in the next section, which in any case is close to unity for 
sufficiently large $f_\mu$.

  
We can also define a fraction 
$f_{\rm vir}$, similarly to $f_{\mu}$, replacing 
$\Delta \Omega_{\mu} \rightarrow \Delta \Omega_{\rm vir}$.  It turns out that
the area of halos defined in this way can be larger than the sky area 
($f_{\rm vir} > 1$) if one includes halos of sufficiently low mass. 
However, those halos are unable to
produce large magnification areas and therefore $f_{\mu}$ remains smaller 
than 1 for sufficiently large values of $\mu_{min}$.

For fixed source redshift and cluster ellipticity (in which case $P(z_s)$ and
$P(e_{\varphi})$ are delta functions) we have
\begin{eqnarray}
\frac{df_{\mu}}{dz_l} &=& 
      \frac{D_A^2(z_l)}{H(z_l)}
      \int_{M_{\rm th}}^{\infty} d\ln \Mvir 
      \Delta\Omega_{\mu}(z_l,\Mvir,\mu_{min}) 
      \frac{dn(z_l,\Mvir)}{d\ln \Mvir}\,, \\
\frac{df_{\rm vir}}{dz_l} &=& 
      \frac{D_A^2(z_l)}{H(z_l)}
      \int_{M_{\rm th}}^{\infty} d\ln \Mvir 
      \Delta\Omega_{\rm vir}(z_l,\Mvir) 
      \frac{dn(z_l,\Mvir)}{d\ln \Mvir}\,. 
\end{eqnarray}

\begin{figure}
\resizebox{85mm}{!}{\includegraphics[angle=0]{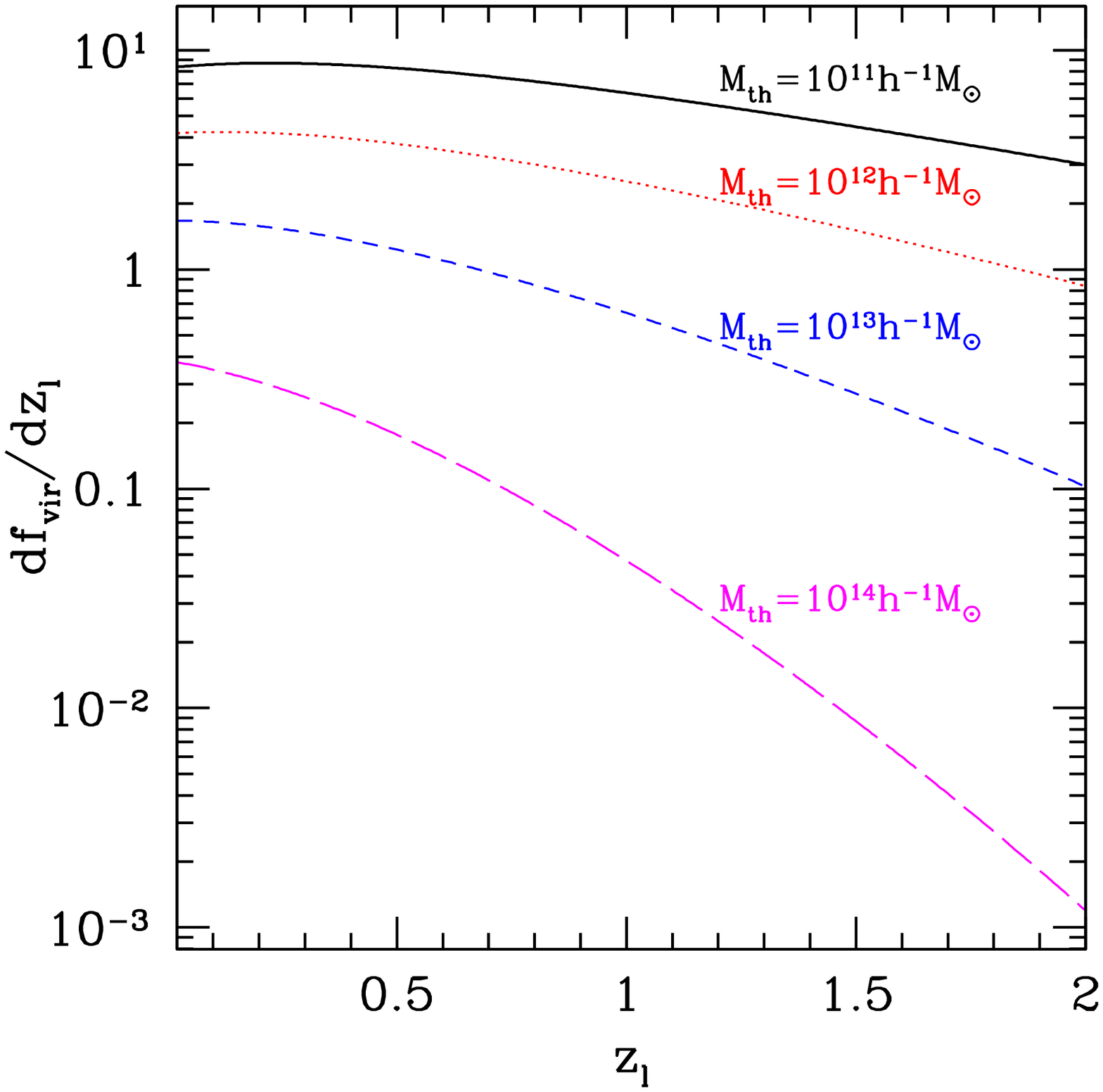}}
\resizebox{85mm}{!}{\includegraphics[angle=0]{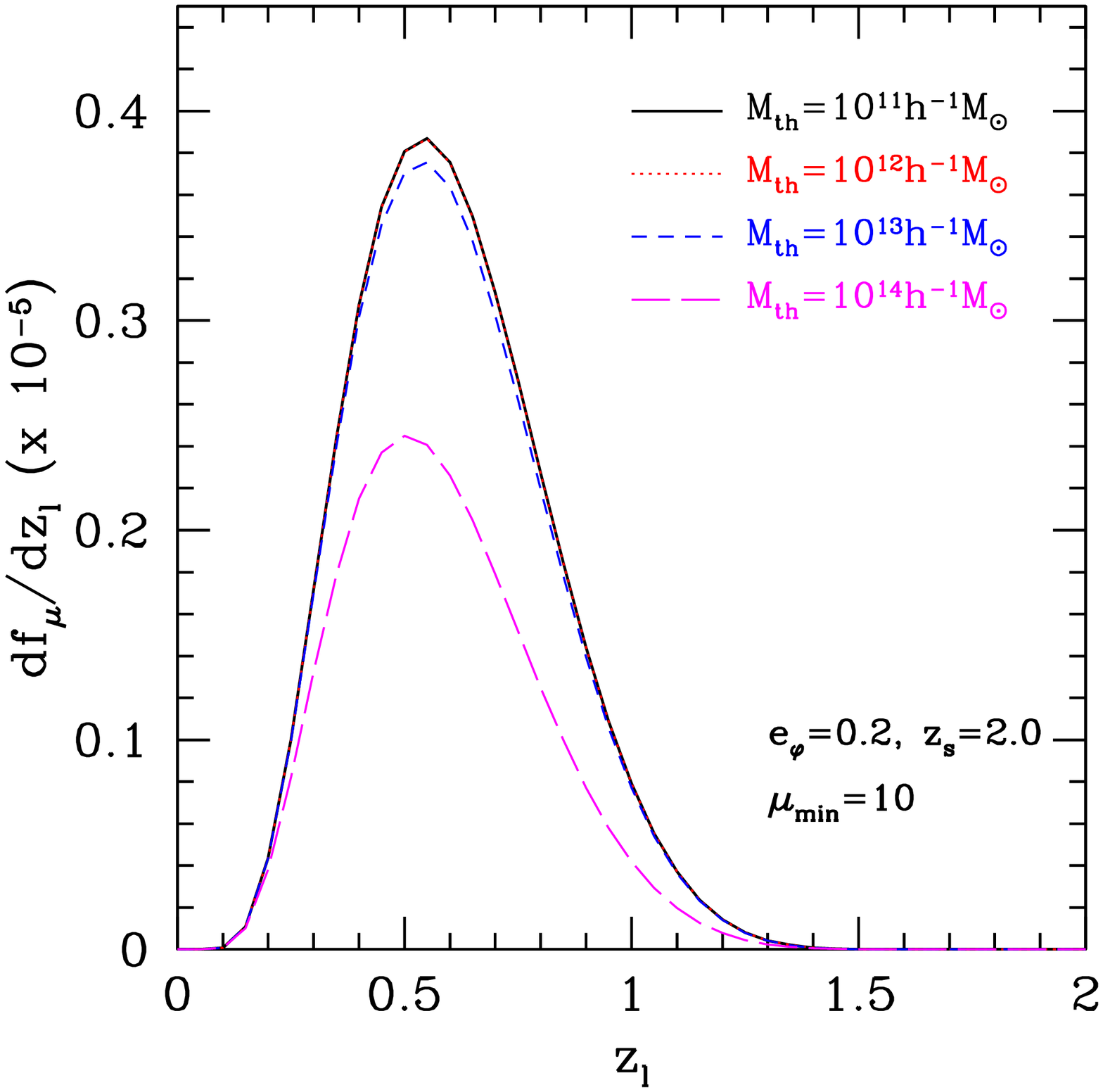}}
  \caption{({\it Left}): Differential sky fraction $df_{\rm vir}/dz_l$ 
as a function of $z_l$ for halos above different mass thresholds 
as indicated. 
({\it Right}): Differential sky fraction  $df_\mu/dz_l$ for halos with 
magnification above $\mu_{min}=10$ and different mass thresholds. 
}
\label{fig:dfdzl_zl}
\end{figure}

In Fig.~\ref{fig:dfdzl_zl} we show the sky fraction 
$df_{\rm vir}/dz_l$ and 
$df_{\mu}/dz_l$ as a function of $z_l$ for fixed 
$z_s=2.0$, $e_{\varphi}=0.2$ and $\mu_{min}=10$. 
The latter fraction does not change much for $M_{\rm th}<10^{12}h^{-1}
M_{\odot}$ since these halos, 
though numerous, are unable to produce large magnifications. 
Very massive halos produce large magnifications but their abundance is 
exponentially suppressed.

\begin{figure}
\resizebox{85mm}{!}{\includegraphics[angle=0]{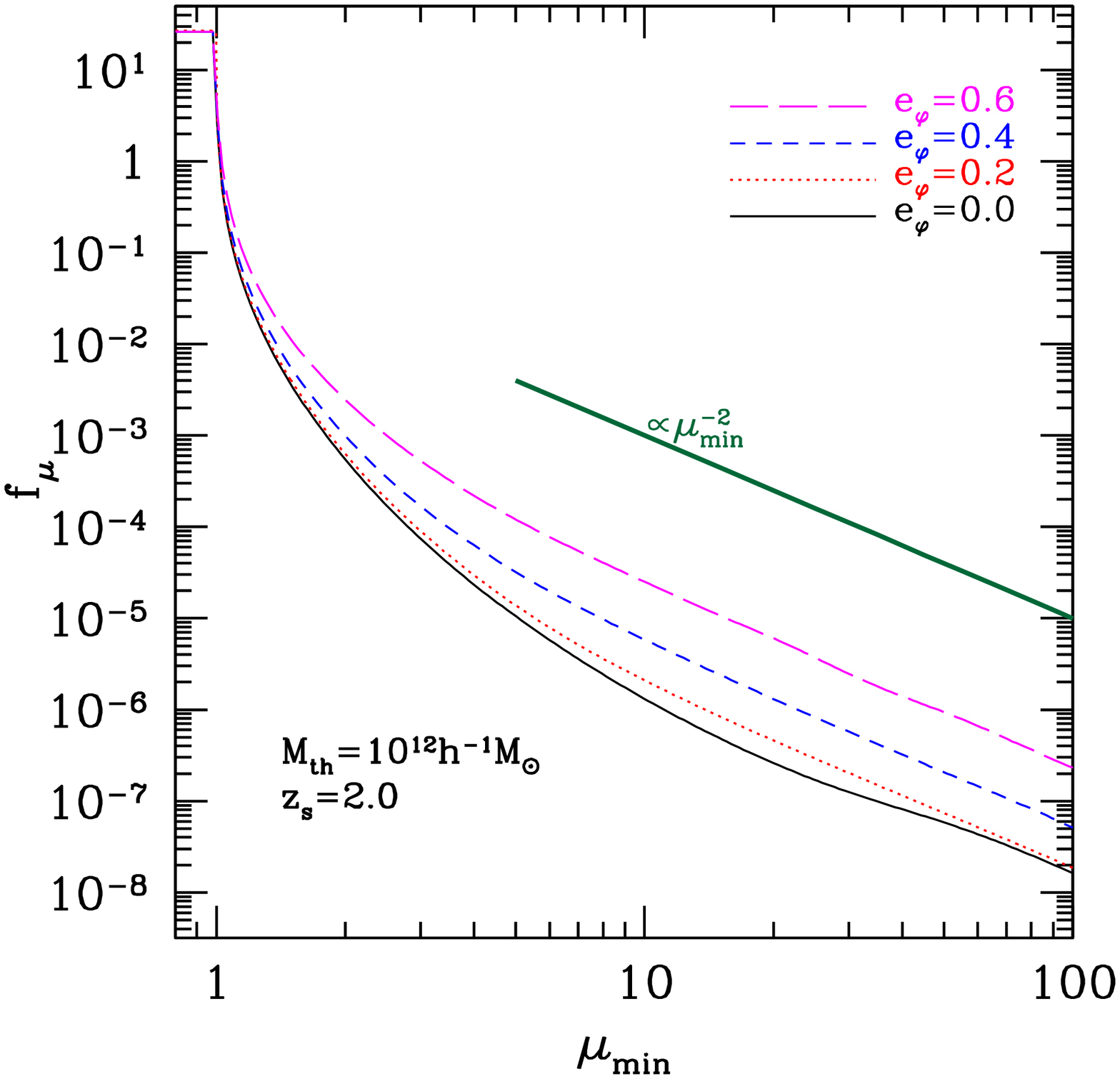}}
\resizebox{85mm}{!}{\includegraphics[angle=0]{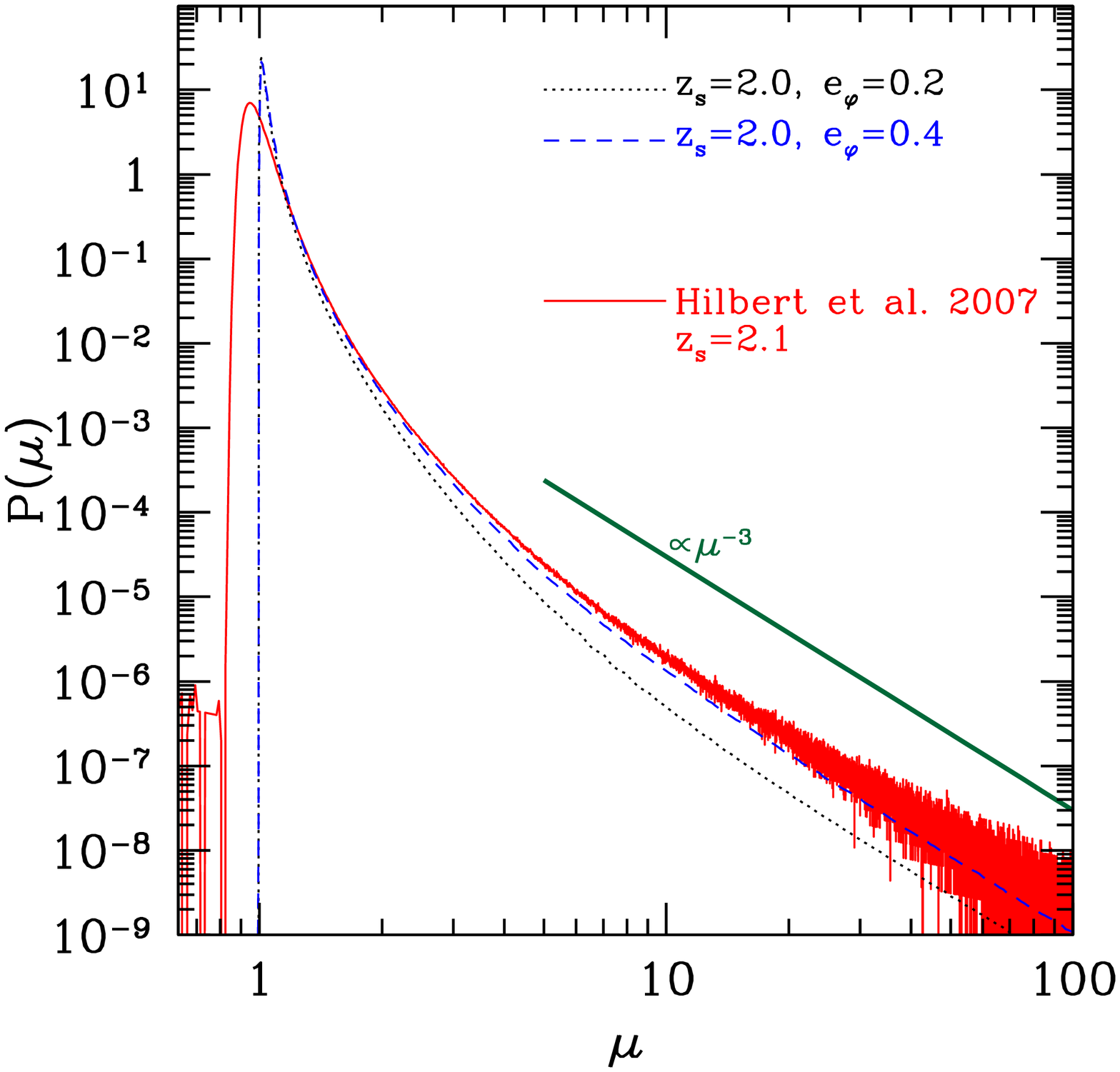}}
  \caption{ ({\it Left}): Sky fraction $f_{\mu}$ with magnification $\mu>\mu_{min}$ as a function of 
$\mu_{min}$ for halos with masses $\Mvir>10^{12}h^{-1} M_{\odot}$ and different ellipticities 
$e_{\varphi}$. 
({\it Right}): Lensing probability $P(\mu)$ to have magnification along the
line of sight up to $z_s=2.0$ from all intervening halos, assumed to have $\ephi=0.2$ (dotted line) 
and $\ephi=0.4$ (dashed line). Also shown is the lensing probability obtained
from N-body simulations (thin solid line) for sources at $z_s=2.1$ \citep{Hiletal07}.   
}
\label{fig:fmu_mumin}
\end{figure}

In the left panel of Fig.~\ref{fig:fmu_mumin} we show the sky fraction $f_{\mu}$ as a
function of $\mu_{min}$ for $M_{\rm th}=10^{12}h^{-1}\Msun$, 
$z_s=2.0$ (left).
The fraction increases sharply as $\mu_{min}$ approaches 1. The
curves with different values of $\ephi$ show that for large
magnifications, $f_\mu$ can be significantly higher
for large ellipticities.  

\subsection{Lensing Probability}  \label{subsec:lensprob}

Given the lensing optical depth $f_{\mu}$ we obtain the lensing probability as
\begin{eqnarray}
P(>\mu)=1-e^{-f_{\mu}}\,.
\end{eqnarray}
Note that $P(>\mu) \approx f_{\mu}$ for $f_\mu \ll 1$ or large magnifications, 
i.e. small optical depths are themselves the lensing probability. 
On the other hand, $P(>\mu) \rightarrow 1$ for $f_\mu \rightarrow \infty$ or
small magnifications. 
Our top cut-off value of $4\thetavir^2$ for $\Delta \Omega_\mu$ implied a 
saturation of $f_\mu \sim 27$ for $\mu \sim 1$. At this value of $f_\mu$, however, 
$P(>\mu)$ is already very close to unit for all practical purposes and 
therefore is not affected significantly by the cut-off.

Note $P(>\mu)$ is the probability that a source galaxy is magnified by 
more than $\mu$ and therefore it can be expressed
in terms of the probability density $P(\mu)$ as
\begin{eqnarray}
P(>\mu)= \int_{\mu}^{\infty} P(\mu) d\mu \,.
\end{eqnarray}
Alternatively $P(\mu)$ can be obtained by 
$P(\mu)=-dP(>\mu)/d\mu$. This statistics can then be incorporated 
in various applications to estimate the effects of lensing magnification.
An interesting property of the lensing probability in the source plane is 
that  $P(>\mu) \propto 1/\mu^2$ and therefore 
$P(\mu) \propto 1/\mu^3$ for $\mu \gg 1$, as can be shown in particular 
cases and argued to be true in general \citep{SchEhlFal92}.

In the right panel of Fig.~\ref{fig:fmu_mumin}, we show $P(\mu)$ 
for $z_s=2.0$ and $\ephi=0.2,0.4$. We also show the magnification probability for sources at $z_s=2.1$
from \cite{Hiletal07}, obtained by ray-tracing high-resolution N-body simulations.
 Notice that, for $\mu \gg 1$, 
$P(>\mu) \sim f_\mu \propto \mu^{-2}$ as shown in the left panel of 
Fig.~\ref{fig:fmu_mumin}, and it follows that $P(\mu) \propto \mu^{-3}$ 
as expected.
Our semi-analytical estimate of $P(\mu)$ with $\ephi=0.4$ compares relatively 
well with the probability obtained from ray-tracing on 
N-Body simulations \citep{Hiletal07, PacScoCha08} at large magnifications.

Even though $\ephi=0.2$ is in better agreement with the average values of 
halo ellipticities \citep{JinSut02}, the fact that our model does 
not include halo substructure 
and other realistic effects such as other halos along the line of sight that enhance magnification appears to be incorporated by artificially increasing the halo ellipticity to $\ephi=0.4$.  
In addition, the semi-analytical model underestimates   
$P(\mu)$ compared to the simulation results at low magnifications 
($\mu \lesssim 1.0$).
That is expected since our model only includes a one-halo term and 
does not properly account for the effects of voids, where highly 
de-magnified regions sit. That feature is well captured
by simulations and can potentially be obtained in an improved version of 
our model that includes compensated halos, a two-halo term \citep{CooShe02} 
as well as halo substructure \citep{Menetal05}. 
In our estimations of lensing results on high-redshift galaxy populations, we 
use $P(\mu)$ from N-body simulations \citep{Hiletal07}.

Finally, note that $P(\mu)$ defined above {\it averages} over clusters in the line 
of sight up to the source redshift $z_s$. When we are interested in the effects
within a specific cluster solid angle, we must use only the region around the
cluster to define the lensing probability. For instance, if we want to estimate 
lensing effects within the virial radius of a specific cluster, we use a probability 
defined simply as $P(>\mu)=\Delta \Omega_\mu/\Delta \Omega_{\rm vir}$, where these 
quantities are calculated from our semi-analytical model described above.

\section{Observable Effects of Lensing Magnification} \label{sec:results}

\subsection{Counts of Submillimeter Galaxies} \label{subsec:SMG}

Submillimeter galaxies (SMGs) are dusty galaxies, typically optically obscured but visible in wavelengths
below $1$mm due to the thermal spectrum re-emitted by the dust grains after absorbing radiation from the star-forming regions.
The simplest prescription for the Spectral Energy Distribution (SED) of 
SMGs is it is 
given by a blackbody spectrum $B_\nu$ at a high temperature $T_{\rm gal}$, modified by 
an emissivity $\epsilon \propto \nu^{\beta}$
\begin{eqnarray}
S_{\nu}^{\rm SED}&=&\epsilon B_{\nu} \propto \nu^{3+\beta}\left\{\exp\left[\frac{h\nu}{k_{\rm B} T_{\rm gal}}\right]-1\right\}^{-1} 
                                 \propto  \nu^{3+\beta}\left\{\exp\left[\frac{(\nu/{\rm GHz})}{21(T_{\rm gal}/{\rm K})}\right]-1\right\}^{-1} \,.
\end{eqnarray}

Apart from an overall normalization factor, the 2 free parameters $\beta$ and $T_{\rm gal}$
as well as the galaxy redshift $z_s$ must be known.
Typical fits to observed galaxies give high redshifts of $z_s\sim 1-6$, values of $\beta$ in 
the range $0.5-2$ and temperatures of $\sim 15-40$K.
Specifically, in our fiducial prescription we assume that all galaxies are 
at $z_s=2$ (though we also show some results for $z_s=3.0$ and $4.0$), and take
$T_{\rm gal}=30$K and $\beta=0.7$.
In the Rayleigh-Jeans regime of low frequencies, the temperature is unimportant
and $S_{\nu}^{\rm SED} \propto \nu^{2+\beta}$. The value of $\beta=0.7$ is consistent
with the spectral index $\alpha=2+\beta$ used by \citet{KnoHolChu04, WhiMaj04}.
Using the SED prescription one can scale fluxes at one 
observed frequency $\nu$ to another frequency of interest $\nu^\prime$, 
e.g. in the SZ flux range. 
As we will see, $\beta=0.7$ properly accounts for extrapolated fluxes in the 
empirically motivated model of \cite{Lagache2004} for the SED of SMGs at 
different frequencies.

The count distribution of SMGs have been measured at various wavelengths, 
including at $250$, $350$ and $500\mu$m by BLAST \citep{Devetal09}, 
at $850\mu$m  
by SCUBA \citep{Copetal06}, at $1100\mu$m by BOLOCAM \citep{Lauetal05} and AzTEC \citep{Peretal08}
and at $1200\mu$m by MAMBO \citep{Greetal04}. See \cite{PeaKha09} for a recent compilation of 
these measurements, a model for fitting them and prospects for future 
measurements.

In our standard prescription we take the differential counts $dn/dS_\nu$ 
of the submillimeter galaxy population as measured by the 
Balloon-borne Large Aperture Submillimeter Telescope (BLAST) at
$\lambda=500 \mu$m ($\nu\sim 600$GHz), and scale various quantities of
interest to other frequencies according to the SED prescription.
When studying the effects of lensing on the properties of SMGs, we also 
consider the example of a Schechter model, in which the exponential
steepness of the  counts make the lensing effects more dramatic.

\subsubsection{Magnification and number counts}
As a result of photon and energy conservation, the surface brightness of 
galaxy sources, defined as the flux per unit solid angle, is conserved by 
gravitational lensing.
Since magnification, by definition, increases the solid angle $\Omega$ 
of sources by a factor $\mu$, it has to also increase their intrinsic 
flux $S$, effectively lowering the observed threshold and increasing 
the number of sources available
\begin{eqnarray}
S&\rightarrow& \Sobs=\mu S\,, \\
d\Omega&\rightarrow& d\Omegaobs=\mu d\Omega \,.
\end{eqnarray}

As a result, the intrinsic number density of a source population 
is modified by lensing magnification.
For a given magnification $\mu$, the intrinsic 
differential number density $dn/dS$ is modified as 
(see e.g. Eq.~(11) in \cite{RefLoe97}) 
\begin{eqnarray}
\frac{dn}{dS} \rightarrow \frac{1}{\mu^2} \frac{dn}{dS}\left(\frac{\Sobs}{\mu}\right) \,.
\end{eqnarray}

The $1/\mu^2$ factor comes from transforming the angle differential 
$d \Omega$ (implicit in the above equation) and the flux differential
$dS$ into their observed counterparts. The additional change is from 
the fact that the observed flux $\Sobs$ corresponds to a true flux of
$S=\Sobs/\mu$. 
The intrinsic differential number density can be interpreted as the 
probability density $P(S)$ of having intrinsic flux $S$ from a random 
source. 
Likewise, the observed modified 
distribution represents the probability density $P(\Sobs|\mu)$ of having 
an observed flux $\Sobs$ given a magnification $\mu$. 
In order to obtain the probability 
$P(\Sobs)$ of having $\Sobs$ irrespective of magnification, we must 
multiply by $P(\mu)$ and integrate over all values of $\mu$. 

Given the differential number density, a number of quantities are of
interest, such as the cumulative number density, the cumulative number 
counts $N(>S)$ 
in a solid angle $\Delta \Omega$, the average total flux of the 
galaxy population $(\barS)^{\rm gal}$ and its associated
Poisson variance $(\barSvar)^{\rm gal}$
\begin{eqnarray}
n(>S)&=&\int_S \frac{dn}{dS^\prime} dS^\prime \,,\\
N(>S)&=&\Delta \Omega \ n(>S) \,, \label{eq:int_counts}\\
\barS&=&\Delta \Omega\int S \frac{dn}{dS} dS \,, \label{eq:int_aveS}\\
\barSvar&=&\Delta \Omega\int S^2 \frac{dn}{dS}dS\,. \label{eq:int_aveS2}
\end{eqnarray}

Ignoring the clustering of the galaxy population, which adds extra noise to the
average flux, the expected fluctuation in the average flux of galaxies 
within solid angle $\Delta \Omega$ due to Poisson noise in the counts is given by
\begin{eqnarray}
\sigma^{\rm gal}(S)=\sqrt{\barSvar }\,.
\end{eqnarray}

The average flux in one frequency $\nu$ can be scaled
to $\nu^\prime$ given a SED prescription $S_\nu^{\rm SED}$
\begin{eqnarray}
\barS_{\nu^\prime}=
             \Delta \Omega\int S_{\nu^\prime} \frac{dn}{dS_{\nu^\prime}} dS_{\nu^\prime} 
             \sim \frac{S_{\nu^{\prime}}^{\rm SED}}{S_\nu^{\rm SED}} 
                  \Delta \Omega\int S_\nu \frac{dn}{dS_\nu} dS_\nu  
             = \frac{S_{\nu^{\prime}}^{\rm SED}}{S_\nu^{\rm SED}} \ \barS_{\nu}\,,
\end{eqnarray}
and a similar rescaling can be applied to $\sigma^{\rm gal}(S_\nu)$.
Lensing magnification modifies these various quantities to their observed 
values according to the effect of a given magnification $\mu$ and the 
lensing probability $P(\mu)$ of actually having that magnification.
Since we define $P(\mu)$ in the {\it source} plane, we may identify 
angular averages in this plane with averages over $P(\mu)$. For a variable $X$,
we have
\bea
\frac{1}{\Delta \Omega}\int d\Omega X =  \int d\mu P(\mu) X\,.
\eea
This implies 
\bea
\langle \mu \rangle = \int d\mu P(\mu) \mu = \frac{\Delta \Omegaobs}{\Delta \Omega}\,.
\eea
Using the relation between averages in the observed (image) and intrinsic
(source) plane, i.e. 
\bea
\frac{1}{\Delta \Omegaobs}\int d\Omegaobs X =\frac{1}{\langle \mu \rangle \Delta \Omega}\int \mu d\Omega X = \frac{1}{\langle \mu \rangle} \int d\mu \ \mu \ P(\mu) X\,,
\eea
and the fact that observations can only average in the image plane, we 
obtain the observed lensed quantities in terms of their unlensed counterparts
\begin{eqnarray}
\frac{d\nobs(\Sobs)}{d\Sobs} &=& \frac{1}{\langle \mu \rangle}\int d\mu \frac{P(\mu)}{\mu} \frac{dn}{dS}\left(\frac{\Sobs}{\mu}\right) \,, \label{eq:dnobsdSobs} \\
\nobs(>\Sobs) &=& \frac{1}{\langle \mu \rangle}\int d\mu P(\mu) n\left(>\frac{\Sobs}{\mu}\right) \,, \\ 
\Nobs(>\Sobs) &=& \frac{1}{\langle\mu\rangle}\int d\mu P(\mu) N\left(>\frac{\Sobs}{\mu}\right) \,, \label{eq:lens_N}\\
\barS_{\rm obs} &=& \barS\,, \label{eq:lens_aveS}\\
\barSvar_{\rm obs} &=& \frac{\langle\mu^2\rangle}{\langle\mu\rangle} \barSvar\,,\label{eq:lens_aveS2}
\end{eqnarray}
where 
\bea
\langle \mu^\alpha \rangle = \int d\mu \ \mu^\alpha P(\mu)\,.
\eea

When considering effects of all halos in the line of sight of source galaxies at 
large sky patches, we use the $P(\mu)$ defined
in \S \ref{subsec:lensprob}. 
When considering the effect of the source population on a single cluster, we use 
$P(\mu)$ defined within the cluster radius. 
Even though an intrinsic solid angle $\Delta \Omega_{\rm int}$ corresponds 
to an observed solid angle 
$\Delta \Omegaobs=\langle \mu \rangle \Delta \Omega_{\rm int}$, 
here we assume that the angle implicit in Eqs.~(\ref{eq:lens_N}-\ref{eq:lens_aveS2}) is 
simply fixed by the observation, i.e. $\Delta \Omegaobs=\Delta \Omega$, 
where $\Delta \Omega$ was used to define unlensed 
quantities Eqs.~(\ref{eq:int_counts}-\ref{eq:int_aveS2}). 
Notice that Eq.~(\ref{eq:lens_aveS}) reflects the conservation of total surface 
brightness by lensing and along with Eq.~(\ref{eq:lens_aveS2}), is only valid 
when these quantities are integrated over all fluxes. If we introduce lower or higher
limits of integration, we need to integrate over $P(\mu)$ as for the other quantities.
In that case Eq.~(\ref{eq:lens_aveS}) is no longer true, i.e. the 
lensed and intrinsic brightness differ.
 
\cite{PacScoCha08} considered the lensing effect on the distribution of 
SMGs detected at $850$ $\mu$m. 
However they did not include the factors of $1/\langle\mu\rangle$ and 
$1/\mu$ in the expression for the differential counts Eq.~(\ref{eq:dnobsdSobs}). 
That caused an overestimation of the magnification effects, 
which are highly sensitive to the steepness of the differential
counts.

If the intrinsic differential  number density can be parameterized
as a power law with index $\alpha$ {\it near} the flux threshold,
\begin{eqnarray}
\frac{dn(S)}{dS} \propto S^{-\alpha} \,,
\end{eqnarray}
then the above expressions become simply
their intrinsic counterparts rescaled by moments of the magnification
probability $P(\mu)$, e.g.
\begin{eqnarray}
\frac{d\nobs(\Sobs)}{d\Sobs} &=& 
            \frac{\langle \mu^{\alpha-1}\rangle}{\langle \mu \rangle}
            \frac{dn(\Sobs)}{dS} \,, \label{eq:powlaw}
\end{eqnarray}
and similarly for $\nobs(>\Sobs)$ and other observables of interest. 
Notice that since $P(\mu) \propto 1/\mu^3$ as $\mu \rightarrow \infty$, the
observed quantities diverge for large enough values of $\alpha$, unless we 
impose an upper integration limit defining a maximum magnification or
a cut-off in the counts distribution.
In practice there is always a maximum magnification imposed, for instance, 
by the size of the source galaxies. Moreover,
a constant power law index does not hold for counts over all fluxes. 
For a non-evolving population in Euclidean flat space $\alpha=2.5$; 
larger/smaller values of $\alpha$ indicate a 
luminosity or density that is increasing/decreasing in time, i.e. non-trivial
evolution of the galaxy population.

\subsubsection{Number Counts from BLAST}

\begin{figure}
\resizebox{88mm}{!}{\includegraphics[angle=0]{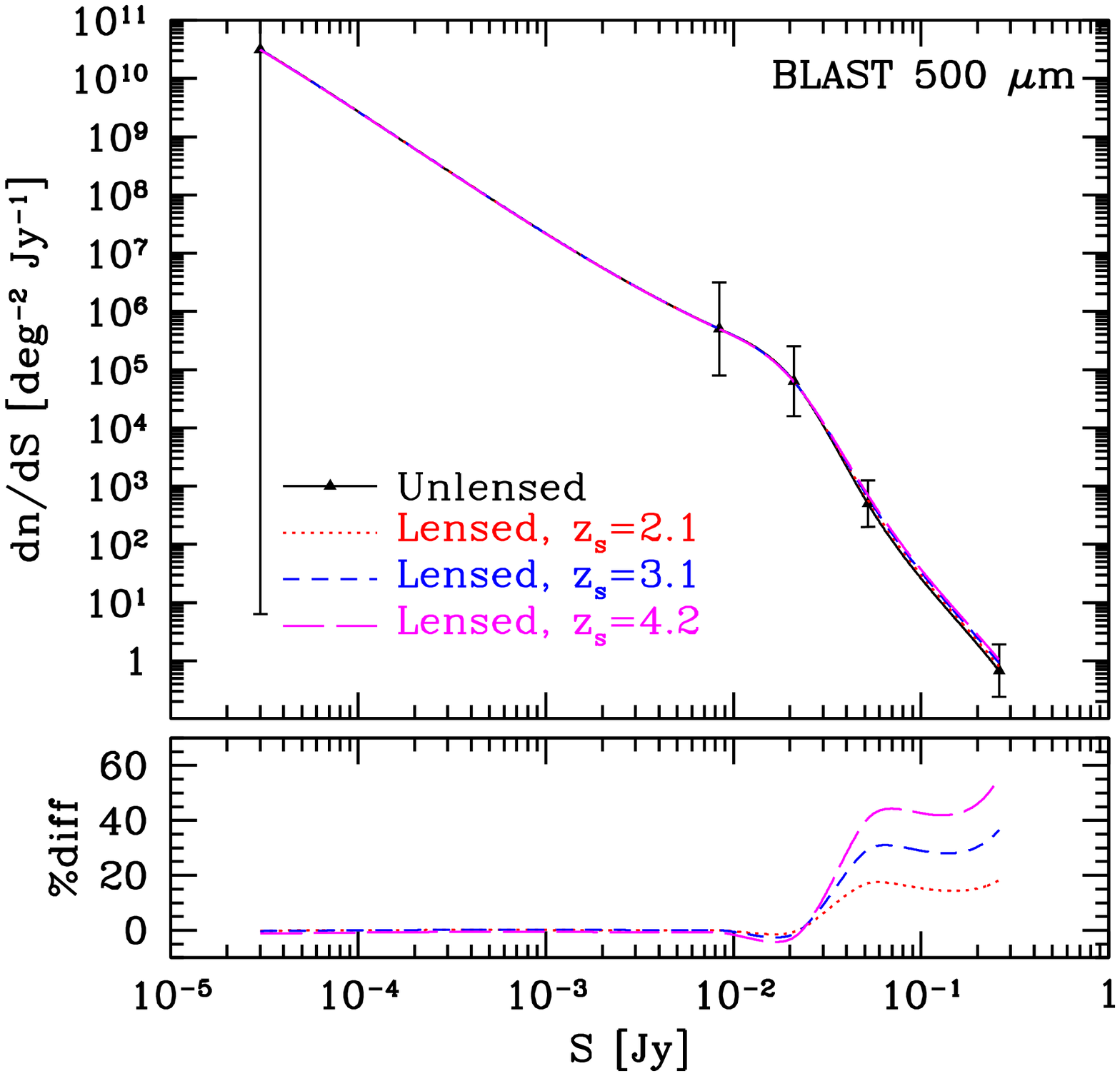}}
\resizebox{88mm}{!}{\includegraphics[angle=0]{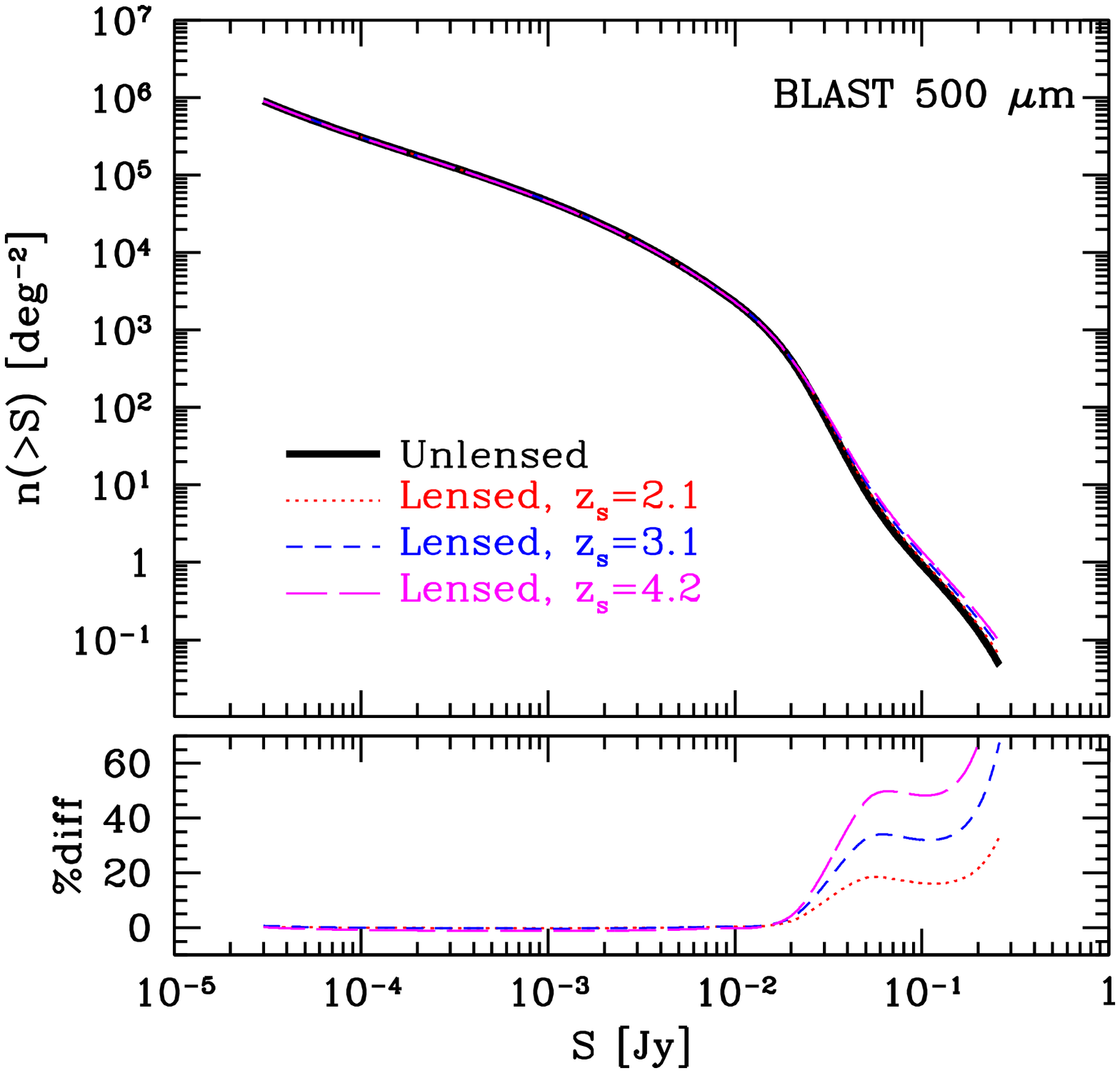}}
  \caption{ ({\it Left}): Intrinsic differential number density $dn/dS$ from BLAST 
$500 \mu$m data (points and solid black line) as well as observed lensed 
counts $d\nobs/d\Sobs$ at different source redshifts (dotted and dashed lines). 
Here the observed BLAST data is assumed to be the true underlying distribution 
at different redshifts, which is then lensed by intervening halos.
({\it Right}): Cumulative number density for the same unlensed/lensed BLAST distribution. 
}
\label{fig:dNdS_BLAST_500}
\end{figure}

In Fig.~\ref{fig:dNdS_BLAST_500} we illustrate lensing effects on the
distribution of SMGs for the BLAST $500 \mu$m data. 
We show the {\it measured} BLAST distributions, which we take to be
intrinsic as opposed to observed for illustrative purposes.
Shown in  Fig.~\ref{fig:dNdS_BLAST_500}
are the differential and cumulative number density, 
as well as the corresponding lensed distributions, 
assuming sources at different redshifts.
We spline interpolate the original BLAST points to obtain a smooth 
distribution at arbitrary values of flux density. 

Obviously, the measured distributions already have lensing effects
convolved, which ideally one would try to deconvolve to obtain the
true distributions. 
If the lensing effects are relatively small, as in this case, the 
above assumption is self-consistent since the observed distribution is in 
fact already close to the true one. 
The data indicates that the observed differential counts have a index 
$\alpha\sim 2$ until $S\sim 10$~mJy and $\alpha\sim4$ beyond this flux. 
Notice that
for $\alpha>3$ the differential number density increases significantly 
(see Eq.~\ref{eq:powlaw}) and the relative difference between
the lensed and unlensed curves increases at the bright end by up to $\sim60\%$.
The empirically motivated model of \cite{Lagache2004}
predicts a distribution at $500\mu$m that is 
quite close to the one measured by BLAST, suggesting that 
the observed slopes are already close to the true ones.
However, in order for the total brightness inferred from the BLAST points 
not to exceed the Cosmic Infrared Background (CIB) 
\citep{Dwek98, Fixen98, Smail97, Hughes98}, it is necessary to impose a 
sharp cut-off flux of $S_{\rm cut}=4.6$~mJy \citep{Devetal09}.
In \S~\ref{subsec:schechter} we consider a case that does not require such cut-off, 
where fluxes decay at the faint end and have 
much steeper intrinsic slopes at the bright end.

Here we have integrated up to a maximum magnification of 
$\mu_{\rm max}=100$. 
However, the results remain nearly the same even for $\mu_{\rm max}=10$, 
reflecting the rarity of high magnification events displayed in 
Fig.~\ref{fig:fmu_mumin}. 
Therefore, our results are insensitive to a 
magnification cut-off introduced by finite size of 
source galaxies (e.g. \cite{TakHam03}). The actual value of $\mu_{\rm max}$ 
depends on the galaxy population considered and for submillimeter 
galaxies is estimated to be $\mu_{\rm max}\sim 10-30$ \citep{Peretal02}. 
However, as discussed in \cite{PacScoCha08}, there are still large 
uncertainties on the sizes of SMGs, and there is one SMG 
lensed by the Abell 2218 cluster, whose magnification is 
estimated to be $\mu\sim 45$ \citep{Kneetal04}.


\subsubsection{Schechter-like distribution} \label{subsec:schechter}

\begin{figure*}
\resizebox{88mm}{!}{\includegraphics[angle=0]{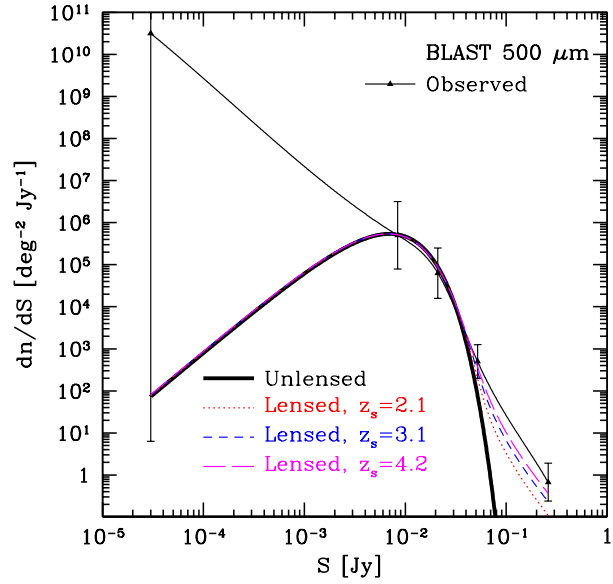}}
  \caption{ Intrinsic differential density $dn/dS$ from BLAST 
$500 \mu$m data (points and thin solid like) as well as observed lensed 
counts $d\nobs/d\Sobs$ at different source redshifts (dotted and dashed lines). 
Here the true underlying distribution is assumed to be of a Schechter type 
(thick solid line), which is then
lensed by intervening halos into the observed BLAST distribution. 
The underlying distribution fits the middle BLAST points and, when lensed,
is roughly consistent with the brightest point. 
}
\label{fig:dNdS_BLAST_500_Schechter}
\end{figure*}

Since the effect of lensing depends on the local slope of the differential counts, 
we expect much larger magnifications if the intrinsic distribution is sufficiently
steep. For illustrative purposes now we assume that the intrinsic counts 
distribution is of a Schechter type \citep{Sch76}
\begin{eqnarray}
\frac{dn(S)}{dS} &=& \left(\frac{N}{S^\prime}\right)
                     \left(\frac{S}{S^\prime}\right)^{\gamma} 
                      e^{-S/S^\prime} \,,
\end{eqnarray}
such that, after lensing, it becomes close to the {\it measured} BLAST distribution.
Beyond the turnaround
flux $S^\prime$, this function becomes extremely steep and 
lensing is very effective.
As an example we fix parameter values for this distribution as
$\gamma=2$, $N=3.5\times10^3$, $S^\prime=3.5$~mJy, 
and show the lensed distributions
 in Fig.~(\ref{fig:dNdS_BLAST_500_Schechter}).
Here the Schechter function fits
intermediate points and decreases fast enough at faint fluxes so as not to 
exceed the CIB flux.
The lensed distribution is consistent with the BLAST data given the error bars
if these galaxies are at redshifts $z_s\gtrsim2$.
In this context, observed counts above $\sim 40$~mJy have all been highly 
magnified from intrinsic lower fluxes. Note the agreement with the brightest point 
can be improved at the expense of a larger disagreement with intermediate points by e.g.
increasing the normalization factor $N$.

In the absence of strong theoretical motivation for the intrinsic distribution,
and given the relatively large error bars on the faint end, we did not attempt 
to fit any particular model to the observed counts.
However, given the two extreme cases considered here, it  
can be seen that the data can be fit by a variety of models, with very 
different implications for the evolution of the luminosity function of these 
galaxies.


\begin{figure*}
  \resizebox{98mm}{!}{\includegraphics[angle=0]{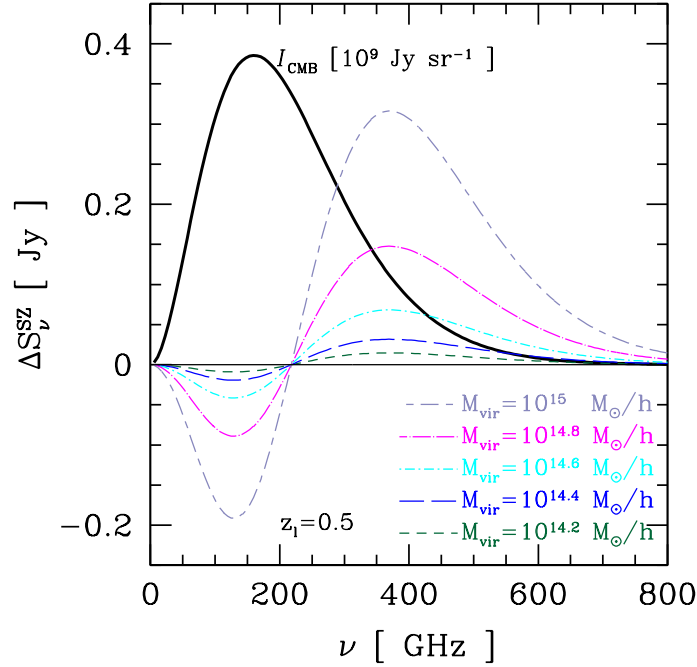}}
  \caption{Blackbody intensity $I_{\rm CMB}$ (thick solid line) and SZ flux 
densities $\Delta S_{\nu}^{\rm SZ}$ for halos at $z_l=0.5$ and virial masses 
ranging from $10^{14.2}$ to $10^{15} h^{-1}M_{\odot}$ (dashed and dot-dashed lines). 
}
\label{fig:SZ}
\end{figure*}

\subsection{SZ Mass Estimation} \label{subsec:SZ}

The CMB has specific 
intensity spectrum $\ICMB$ of a blackbody with temperature $\TCMB \sim 2.725 K$. 
In terms of
$x=h\nu/\KB\TCMB=\nu/(56.78$GHz) and 
$I_0= 2h/c^2(\KB\TCMB/h)^3 = 2.699 \times 10^8$Jy sr$^{-1}$ 
we have 
\begin{eqnarray}
\ICMB&=&B_{\nu}=\frac{2h\nu^3}{c^2}\left(e^{h\nu/\KB\TCMB} -1\right)^{-1}
             =\frac{I_0x^3}{e^x-1}\,.
\end{eqnarray}

As CMB photons cross the cluster hot gas and interact with its high temperature 
electrons via inverse Compton scattering, the outgoing photons gain energy, shifting their 
occupation number, temperature and intensity.

The change in specific intensity can be computed by solving the full
relativistic 
Boltzmann equation and expanding in powers of $\theta_e=(\KB
T_e/m_ec^2)$ \citep{Ito98} to obtain
\begin{eqnarray}
\frac{\Delta I_{\nu}}{I_0}= f(x,T_e)g(x)y\,,
\end{eqnarray}
where $T_e$ is the electron temperature, 
$g(x)=x^4e^x/(e^x-1)^2$, the Compton $y$-parameter is defined as
\begin{eqnarray}
y=\sigma_T \theta_e \int n_e dl = \frac{\KB\sigma_{\rm T}}{m_ec^2} \int n_e(l)T_e(l)dl\,,
\end{eqnarray}
and
$f(x,T_e)
=f_0(x)=x(e^x+1)/(e^x-1)-4$, neglecting terms of higher order in 
$\theta_e$, which are negligible in the non-relativistic regime. 
Since $I_{\nu}=dS_{\nu}/d\Omega$, the change in flux $\Delta S_{\nu}$ through the 
cluster is computed by integrating over the cluster solid angle $d\Omega=dA/d_A^2$
\begin{eqnarray}
\Delta S_{\nu}^{\rm SZ}= \int \Delta I_{\nu} d\Omega =I_0 f(x,T_e)g(x) Y\,, \label{eq:DeltaS_SZ}
\end{eqnarray}
with the integrated Compton $Y$-parameter given by 
\begin{eqnarray}
Y=\int y d\Omega
 =\frac{\KB\sigma_{\rm T}}{m_ec^2} \int n_e T_e dl d\Omega
 =\frac{1}{d_A^2}\frac{\KB\sigma_{\rm T}}{m_ec^2} \int n_e T_e dV\,,
\end{eqnarray}
and the cluster volume element is $dV=dAdl=d_A^2 d\Omega dl$. 
One can model electron density $n_e$ and temperature $T_e$ 
profiles and compute $Y$ from them. 
Since the total number of electrons in the cluster $N_e =\int n_e dV$ is 
proportional to the total cluster mass times the gas fraction $f_{\rm
  gas}$, one expects the 
scaling $\int n_e T_e dV \propto \Mgas \Tgas$. 
Further considerations for the case of a gas in virial equilibrium 
produce the expected scaling relation
\begin{eqnarray}
Y &\propto& \fgas^{5/3} \frac{E(z)^{2/3}}{d_A^2(z)} M^{5/3}\,.
\end{eqnarray}
We employ fits from hydrodynamic simulations of \citet{Nag06} for this relation given by
\begin{eqnarray}
Y &=& A_{14}\times 10^{-6} E(z)^{2/3} 
            \left(\frac{h^{-1}{\rm Mpc}}{d_A(z)}\right)^2 
            \left(\frac{M}{10^{14}h^{-1}M_{\odot}}\right)^{\alpha_m}\,, 
\end{eqnarray}
Therefore, given the cluster mass $M$ and redshift $z_l$ we can estimate
its integrated $Y$ parameter through the scaling 
relation and the change in flux density follows from Eq.~(\ref{eq:DeltaS_SZ}).

In our results we consider two mass definitions for the scaling relation. 
We use parameter values 
$A_{14}=2.5$, $\alpha_m=5/3$, which 
are appropriate for the mass $\Mvir$ defined within the virial radius. 
These values are
roughly intermediate between adiabatic simulations and those with cooling and 
star formation processes.
Similarly, we take 
$A_{14}=7.0$, $\alpha_m=5/3$, for the mass $M_{2500}$ defined within the radius
$r_{2500}$ where the halo overdensity is $2500$ times the critical density.

In Fig.~\ref{fig:SZ} we show the blackbody spectrum $I_{\rm CMB}$ as well as 
the SZ distortion for halos at $z_l=0.5$ and with different virial masses. Here we employed 
the non-relativistic limit $f(x,T_e)=f_0(x)$ for illustrative purposes. 

\subsubsection{Contamination of the SZ signal}

We now estimate the effect of SMGs and their magnification 
on the SZ signal of clusters. SMGs have typical spectra that peak at $\sim 100 \mu$m, 
but they still have significant flux around $\sim 1$mm, where the SZ effect
becomes important.

Since the SZ effect is measured by subtracting the flux of the CMB photons in 
the line-of-sight of the halo from the average flux of background CMB photons, 
adding an average flux $\barS_{\nu}$ from
background galaxies in principle 
does not affect the signal, since the background is removed by the subtraction.
This remains true even with lensing since the average flux does not change 
(Eq.~\ref{eq:lens_aveS}).  However,
local flux fluctuations $\sigma^{\rm gal}(S_{\nu})$ in the background flux due to 
Poisson noise in galaxy counts cannot be removed. 

The true SZ flux $\Delta S_\nu^{\rm SZ}$ is then modified by
$\sigma^{\rm gal}(S_{\nu})$. These fluctuations are further
increased by lensing magnification to $\sigma^{\rm gal}_{\rm obs}(S_{\nu})$ so that
the $68\%$ range of the observed SZ flux becomes
\begin{eqnarray}
(\Delta S_{\nu}^{\rm SZ})_{\rm obs} = \Delta S_\nu^{\rm SZ} \pm \sigma^{\rm gal}_{\rm obs}(S_{\nu})
\end{eqnarray}

In the absence of lensing effects the flux noise is simply
$\sigma^{\rm gal}_{\rm obs}(S_{\nu})=\sigma^{\rm gal}(S_{\nu})=\sqrt{\barSvar_{\nu}}$ 
where the fluctuation is within the cluster virial radius
\begin{eqnarray}
\barSvar_{\nu} = \Delta \Omega_{\rm vir} \int dS_{\nu} S_{\nu}^2 \frac{dn}{dS_{\nu}}\,.
\end{eqnarray}
 
\begin{figure*}
\resizebox{88mm}{!}{\includegraphics[angle=0]{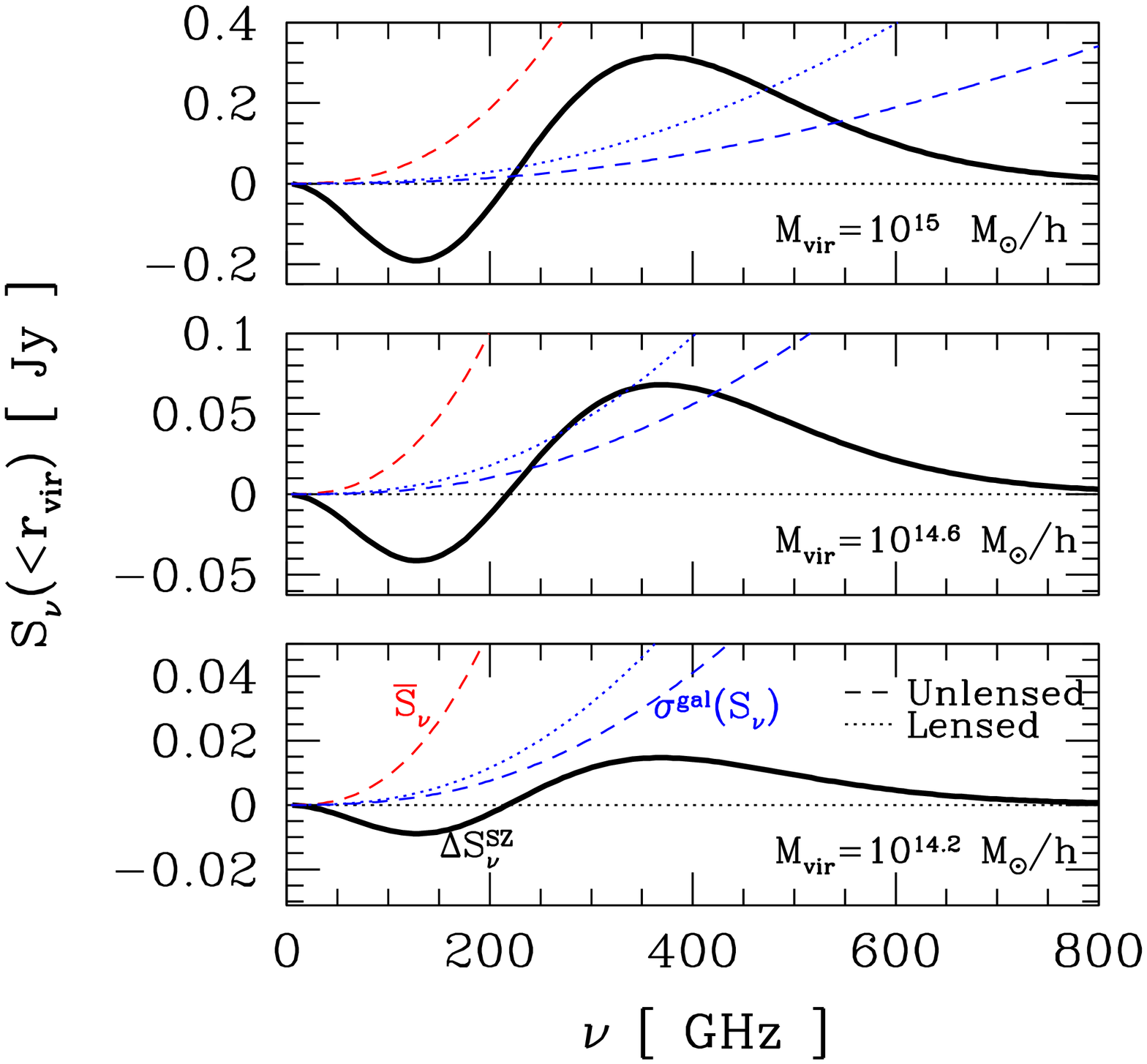}}
\resizebox{88mm}{!}{\includegraphics[angle=0]{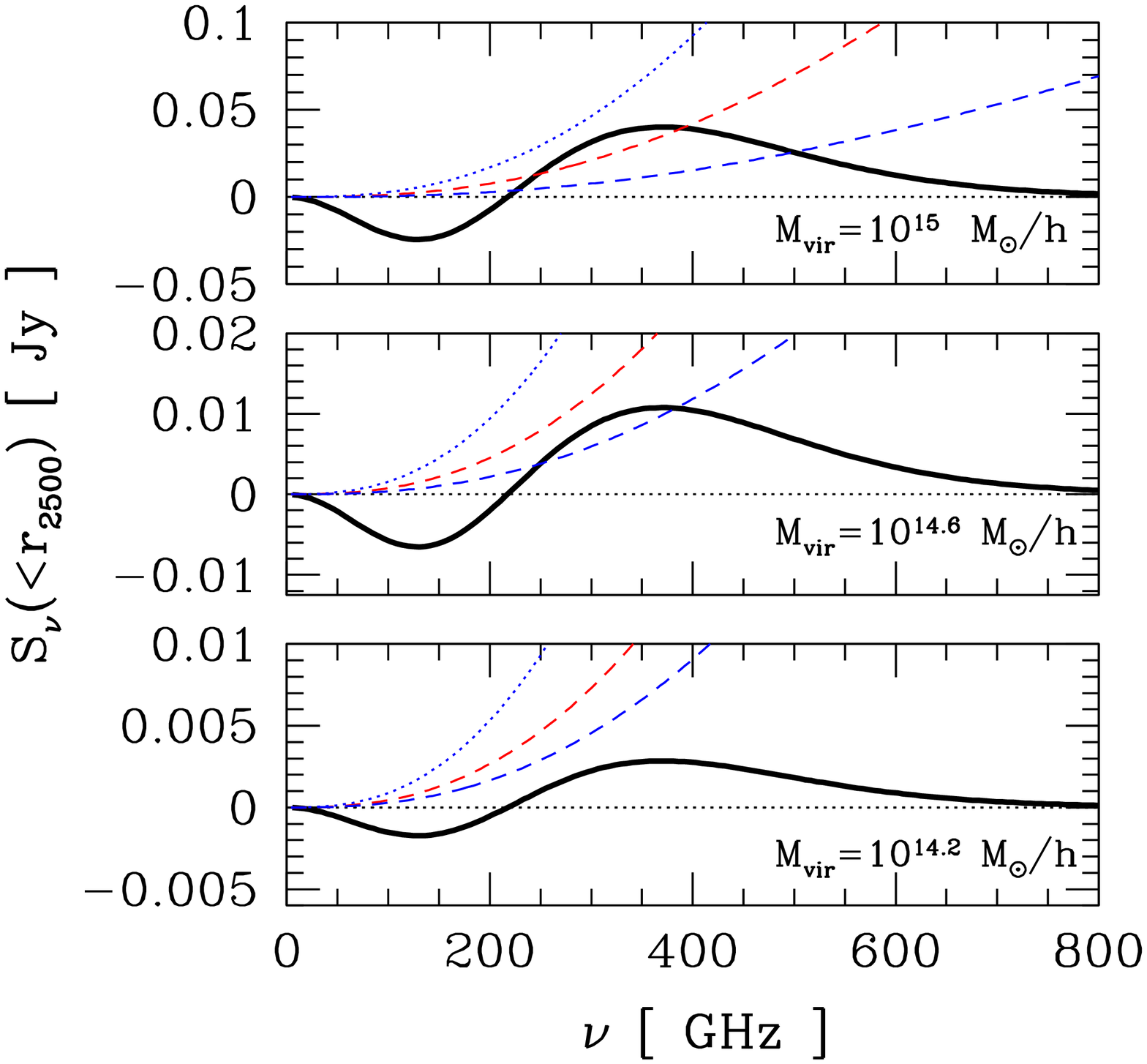}}
  \caption{ SZ flux for clusters of different masses at $z_l=0.5$ 
(solid thick line) and the 
corresponding average background flux $\barS_{\nu}$ 
and noise $\sigma^{\rm gal}(S_{\nu})$ within the cluster radius 
from a population of SMGs at $z_s=2.0$ (dashed lines). 
({\it Left}): Fluxes measured within the virial radius $\rvir$.
({\it Right}): Fluxes measured within $r_{2500}$. Radii and masses are converted
with the prescription of \citet{HuKra03} for a NFW profile.
For clusters of larger masses, the fluctuations from background galaxies
increase, but become a smaller fraction of the total SZ signal. 
Lensing magnification (dotted lines) further enhances 
the flux noise, especially within smaller apertures.
}
\label{fig:SZ_Sobs}
\end{figure*}

In Fig.~\ref{fig:SZ_Sobs}, we show the SZ flux for clusters of various 
masses at $z_l=0.5$ and the contributions of background SMGs. 
Here the average flux and its noise are computed with the BLAST distribution at 
$\lambda=500~\mu$m ($\nu=600$~GHz) and extrapolated to other frequencies using the
SED prescription. 
The average background flux $\barS_{\nu}$ from these
galaxies within the virial radius of the cluster is rather
significant: it is 
larger than the SZ signal. However, as a ``background'' it
can be subtracted. Without lensing, the noise $\sigma^{\rm gal}(S_{\nu})$ in the flux is 
substantially smaller than the background, but non-negligible in
comparison to the SZ signal. 

For clusters of large mass, even though the fluctuations from background 
galaxies increase due to the larger cluster area, these fluctuations
become a smaller fraction of the total SZ signal, and relatively less
important. 
Less massive clusters however are highly affected by the fluctuations of  
SMGs. For clusters of mass $10^{14.6}h^{-1}M_{\odot}$, expected to be above the 
minimum mass detected by upcoming surveys, even the unlensed contamination is 
comparable to the SZ increment. 
The right panels of Fig.~\ref{fig:SZ_Sobs} show the effect of using a 
different aperture in the definition of mass and SMG contamination. 
Note the relative level of contamination remains about the same without lensing.  

In both panels, magnification due to the cluster is rather significant, especially in the
case of SZ fluxes through smaller radii ($r_{2500}$), since those are confined to 
cluster cores of high magnification. The flux noise is magnified by factors of $\sim 2-7$. 
Magnification due to large-scale structure in the line of sight is much smaller -- 
it enhances the noise by only a few percent.

\begin{figure*}
\resizebox{88mm}{!}{\includegraphics[angle=0]{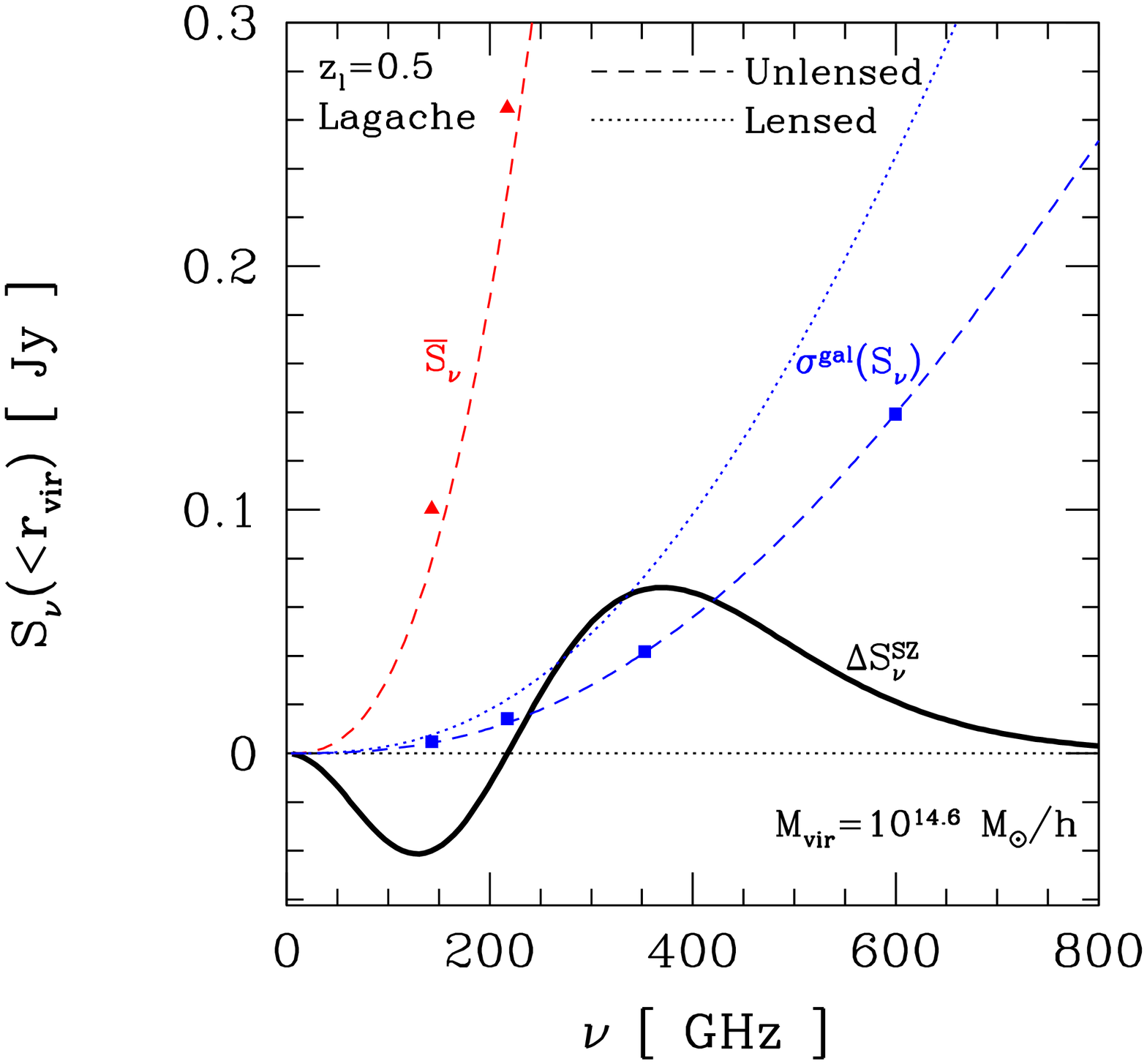}}
\resizebox{88mm}{!}{\includegraphics[angle=0]{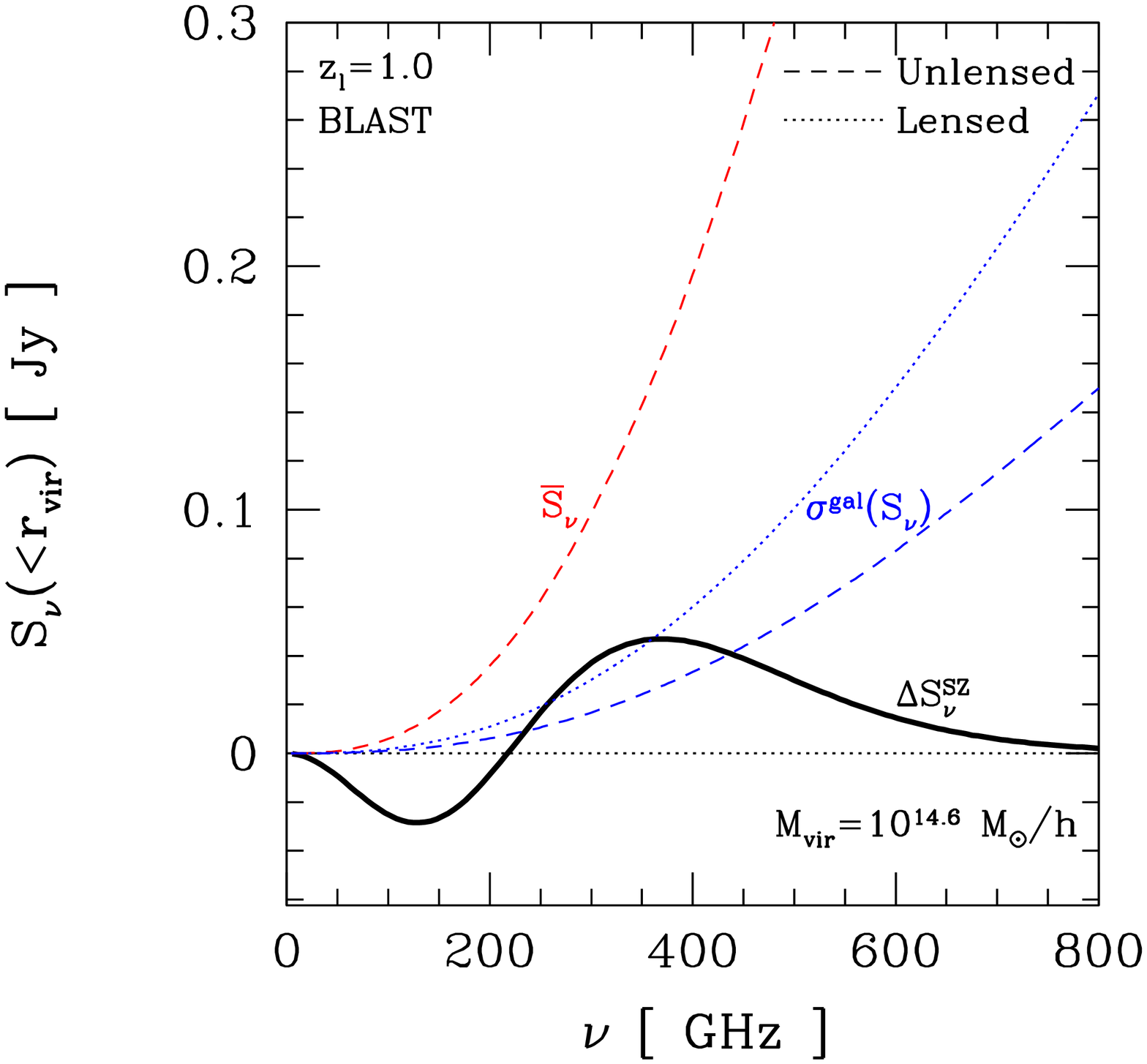}}
  \caption{ SZ flux for clusters with a virial mass of $10^{14.6}
    h^{-1}M_{\odot}$, as in the middle left panel of 
    Fig.~\ref{fig:SZ_Sobs}. 
   ({\it Left}): Instead of using the BLAST data, we employ the Lagache model for estimating 
    contamination by SMGs. 
    The dashed and dotted lines are still computed extrapolating from the $\nu=600$GHz value
    with the SED prescription. Triangle and square dots denote average fluxes and Poisson 
    fluctuations at exact wavelengths $\lambda=500,850,1380,2097\mu$m, and show that the 
    extrapolation agrees well with the model.
   ({\it Right}): Same as middle left panel of Fig.~\ref{fig:SZ_Sobs}, but for cluster redshift 
    $z_l=1.0$. Both $\Delta S^{\rm SZ}_\nu$ and the contribution of SMGs decrease 
    for higher $z_l$, but the contamination level remains roughly the same.
}
\label{fig:SZ_Sobs2}
\end{figure*}

The left panel of Fig.~\ref{fig:SZ_Sobs2} shows results for clusters of
virial mass $10^{14.6} h^{-1}M_{\odot}$, but with the Lagache model 
for the flux from SMGs \citep{Lagache2004}. 
As mentioned earlier, the Lagache model agrees well with the BLAST data at
$\lambda=500\mu$m, so the dotted and dashed lines are nearly identical to those of
the middle left panel of Fig.~\ref{fig:SZ_Sobs}. For the Lagache model, the 
count distributions are available at wavelengths within the SZ range. 
The triangle and square dots shown here are computed directly from the Lagache model at
$\lambda=500,850,1380,2097\mu$m ($\nu=600,350,220,140$GHz). Notice that the SED 
extrapolation from $\lambda=500\mu$m (dotted and dashed lines) fits well the
model points.

The right panel of Fig.~\ref{fig:SZ_Sobs2} shows the results for the BLAST data, but
for a cluster redshift of
$z_l=1.0$. The qualitative results again do not change.  We have
verified this for other masses and redshifts as well. 

\section{Conclusion} \label{sec:conclusion}

We have  implemented a halo model calculation of magnification 
effects of galaxy and cluster halos. We explored the reliability of our model and compared 
it with published results from simulations. The one-halo term computed in the model describes
well the magnification probability at large magnifications, but undersestimates it at 
regions of de-magnification ($\mu<1$). The effects of compensating halos and addition of 
a two-halo term \citep{CooShe02}, as well as the effects of cluster substructure can further 
improve our semi-analytical calculation.  

We applied magnification probability distributions from numerical
simulations (Hilbert et al. 2007) to the counts of distant galaxies. 
Our results differ from those of \cite{PacScoCha08}: our expression differs 
from theirs by two magnification factors, leading to a significantly smaller lensing 
contribution than their estimate. 
For the counts of high redshift SMGs
recently observed, we find that lensing 
leads to enhancements of at least 60 percent at the bright (steep) end of the counts . 
In particular cluster fields with large magnification cross-section, this enhancement can be 
significantly larger.

Galaxy cluster surveys that utilize the SZ effect are susceptible to at least
three kinds of contamination: by radio galaxies associated with the cluster, by high 
redshift SMGs that lie behind the cluster, and by projection effects due to
large scale structure. The latter may result for instance from additional groups and clusters
below the survey detection threshold which end up being projected on observed clusters 
\citep{Haletal07}.
Here we estimated the contribution of SMGs over the frequency range targeted by 
SZ surveys. 
We used the counts from the BLAST survey, extrapolated to lower frequencies, and included 
the effects of lensing magnification from simulations and semi-analytical estimates.

We find that for clusters with masses between $10^{14}$ and $10^{15}$ solar masses, 
there is a significant contamination from SMGs at frequencies higher than 
the SZ null at 220 GHz. This contribution arises from the irreducible variance of the  
Poisson distribution of the SMGs.
It is comparable to the SZ excess at 350 GHz at the low mass end, and is about half 
the SZ excess for a mass of $\Mvir =4 \times 10^{14}$ solar masses 
(it can be either an excess or a deficit for a particular cluster). 
With arcminute sized beams, SMGs are confusion limited, so it is not possible to 
isolate and remove their contribution. 
These results are consistent with earlier studies of the (unlensed) contribution 
estimated from SCUBA sources \citep{WhiMaj04}.
We also find that lensing from the cluster cores significantly enhances this irreducible
noise, especially for SZ fluxes within small appertures containing the cluster critical
curves.

Our results imply that SZ surveys must model the contamination noise of SMGs at the 
SZ null and for channels at higher frequencies. 
The value of the signal measured in such channels is significantly contaminated, even 
for the highest mass clusters due to lensing magnification. 
At frequencies below the SZ null, the SMG contribution is at the few 
percent level for mass ranges of interest. 
So it is less likely to be a problem in cluster detection in the SZ decrement regime, 
but must be included for cosmological measurements, as even a few percent bias in 
the inferred cluster mass can affect derived cosmological parameters due to the 
steepness of the mass function (especially at the high mass end).

\section*{Acknowledgments}

We thank Matthias Bartelmann, Edward Chapin, Anya Chaudhuri, Gary Bernstein, Alex Borisov, 
Jose Diego, Jacek Guzik, Eric Hallman, Wayne Hu, Mike Jarvis, Danica Marsden,
Ravi Sheth, Peter Schneider, Masahiro Takada and Martin White 
for useful discussions and 
Stefan Hilbert for sharing his simulation results. 
We also thank the participants of the DES 
collaboration meeting in Rio de Janeiro and the summer workshops at
the Aspen Center for Physics for fruitful discussions. This work was
supported in part by an NSF-PIRE grant and AST-0607667.

\bibliographystyle{mn2e}
\bibliography{lensing}

\appendix

\section{Lensing by ellipsoidal halos}

The effects of lensing on source images are introduced in Section
3.1. It is useful here to recall basic results for the axially symmetric lens
case before considering the elliptical case. In general the deflection
angle vector $\vec \alpha=(\alpha_1,\alpha_2)$ relating source and image angles 
can be obtained from the convergence field $\kappa(\theta)$ by 
\begin{eqnarray}
\vec \alpha(\vec \theta)=\frac{1}{\pi} \int d^2\theta^{\prime} \kappa(\theta^{\prime}) 
       \frac{\vec \theta - \vec \theta^{\prime}}{|\theta-\theta^{\prime}|^2} \,,
\end{eqnarray}
where $\vec \theta=(\theta_1,\theta_2)$.
If the density profile has axial symmetry, so do all statistical 
quantities derived from it. In particular the lensing potential is 
independent of the position angle with respect to the lens center
$\varphi(\theta_1,\theta_2)=\varphi(\theta)$. Choosing the lens center
as the origin of the coordinate system, the deflection vector
points towards the lens center with magnitude
\begin{eqnarray}
\alpha(\theta)=\frac{1}{\pi \theta} \int 2\pi \theta^{\prime} d\theta^{\prime} \kappa(\theta^{\prime})=\frac{m(\theta)}{\theta}\,,
\end{eqnarray}
where $m(\theta)=M(<\theta)/\pi D_l^2\Sigma_{\rm crit}$ with $M(<\theta)$ 
being the mass within $\theta$.

Given an axially symmetric lensing potential $\varphi(\theta)$, we obtain
the elliptical generalization with major axis along the $\theta_2$ direction 
by replacing $\theta^2=\theta_1^2+\theta_2^2$ by
\begin{eqnarray}
\theta\rightarrow\bar{\theta}=\sqrt{\frac{\theta_1^2}{(1-\ephi)}+\theta_2^2(1-\ephi)}\,.
\end{eqnarray}

Using the derivatives
$\partial\bar{\theta}/\partial  \theta_1$ and 
$\partial\bar{\theta}/\partial  \theta_2$, 
the Cartesian components of the deflection angle are given by
\begin{eqnarray}
\alpha_1&=&\frac{\partial \varphi}{\partial \theta_1}
         = \frac{\theta_1}{(1-\ephi)\bar{\theta}}\tilde{\alpha}(\bar{\theta}) \,, \nonumber \\
\alpha_2&=&\frac{\partial \varphi}{\partial \theta_2}
         = \frac{\theta_2(1-\ephi)}{\bar{\theta}}\tilde{\alpha}(\bar{\theta}) \,,
\end{eqnarray} 
and their derivatives are
\begin{eqnarray}
\frac{\partial \alpha_1}{\partial \theta_1}&=&
          \frac{\tilde{\alpha}(\bar{\theta})}{(1-\ephi)\bar{\theta}} 
          +\frac{\theta_1^2}{(1-\ephi)^2\bar{\theta}^2}
           \left(-\frac{\tilde{\alpha}(\bar{\theta})}{\bar{\theta}} + \frac{d{\tilde{\alpha}}(\bar{\theta})}{d\bar{\theta}}\right)\,, \nonumber \\ 
\frac{\partial \alpha_2}{\partial \theta_2}&=&
         \frac{\tilde{\alpha}(\bar{\theta})(1-\ephi)}{\bar{\theta}} 
          +\frac{\theta_2^2(1-\ephi)^2}{\bar{\theta}^2}
           \left(-\frac{\tilde{\alpha}(\bar{\theta})}{\bar{\theta}} + \frac{d{\tilde{\alpha}}(\bar{\theta})}{d\bar{\theta}}\right)\,, \nonumber \\
\frac{\partial \alpha_1}{\partial \theta_2}&=&\frac{\partial \alpha_2}{\partial \theta_1}=
          \frac{\theta_1\theta_2}{\bar{\theta}^2}
           \left(-\frac{\tilde{\alpha}(\bar{\theta})}{\bar{\theta}} + \frac{d{\tilde{\alpha}}(\bar{\theta})}{d\bar{\theta}}\right)\,. \label{eq:genellip}
\end{eqnarray} 

Here and below tildes denote spherically symmetric quantities, e.g. 
$\tilde{\alpha}(\bar{\theta})=d\varphi/d\bar{\theta}$. 
First, let us check the axially symmetric case by setting $\ephi=0$, in which case
$\bar{\theta}^2=\theta^2=\theta_1^2+\theta_2^2$ and
\begin{eqnarray}
\frac{\partial \tilde{\alpha_i}}{\partial \theta_j}&=&
          \delta_{ij}\frac{\tilde{\alpha}(\bar{\theta})}{\bar{\theta}} 
          +\frac{\theta_i\theta_j}{\bar{\theta}^2}
           \left(-\frac{\tilde{\alpha}(\bar{\theta})}{\bar{\theta}} + \frac{d{\tilde{\alpha}}(\bar{\theta})}{d\bar{\theta}}\right)\,. 
\end{eqnarray}

In this case, the convergence and shear simplify to
\begin{eqnarray}
\tilde{\kappa}&=&\frac{1}{2}\left(\frac{\tilde{\alpha}(\bar{\theta})}{\bar{\theta}} 
               + \frac{d{\tilde{\alpha}}(\bar{\theta})}{d\bar{\theta}}\right)\,, \nonumber \\
\tilde{\gamma}_1&=&\frac{\theta_1^2-\theta_2^2}{2\bar{\theta}^2}
                      \left(-\frac{\tilde{\alpha}(\bar{\theta})}{\bar{\theta}} 
               + \frac{d{\tilde{\alpha}}(\bar{\theta})}{d\bar{\theta}}\right), \ \ \ \tilde{\gamma}_2=\frac{\theta_1\theta_2}{\bar{\theta}^2}
                     \left(-\frac{\tilde{\alpha}(\bar{\theta})}{\bar{\theta}} 
              + \frac{d{\tilde{\alpha}}(\bar{\theta})}{d\bar{\theta}}\right)\,, \nonumber \\ 
|\tilde{\gamma}|&=&-\frac{1}{2}\left(-\frac{\tilde{\alpha}(\bar{\theta})}{\bar{\theta}} 
               + \frac{d{\tilde{\alpha}}(\bar{\theta})}{d\bar{\theta}}\right)\,. 
\end{eqnarray}

Notice that the average value of $\tilde{\kappa}$ within
$\bar{\theta}$ is
\begin{eqnarray}
\bar{\tilde{\kappa}}(\bar{\theta})&=&\frac{1}{\pi \bar{\theta}^2}\int_{0}^{\bar{\theta}}2\pi \theta^{\prime} \tilde{\kappa}(\theta^{\prime})d\theta^{\prime}
             =\frac{\tilde{\alpha}(\bar{\theta})}{\bar{\theta}}\,,
\end{eqnarray}
and therefore it is easy to verify the fact that
\begin{eqnarray}
|\tilde{\gamma}|&=&\frac{1}{2}\left(\frac{\tilde{\alpha}(\bar{\theta})}{\bar{\theta}} 
               - \frac{d{\tilde{\alpha}}(\bar{\theta})}{d\bar{\theta}}\right)
         =\bar{\tilde{\kappa}}-\tilde{\kappa}\,.
\end{eqnarray}

Using the following equalities
\begin{eqnarray}
         \frac{\tilde{\alpha}(\bar{\theta})}{\bar{\theta}}&=&\tilde{\kappa}+|\tilde{\gamma}|\,,  \nonumber \\
\left(-\frac{\tilde{\alpha}(\bar{\theta})}{\bar{\theta}} + \frac{d{\tilde{\alpha}}(\bar{\theta})}{d\bar{\theta}}\right) 
&=&-2|\tilde{\gamma}|\,, 
\end{eqnarray}
in the expressions for general ellipticity Eqs.~(\ref{eq:genellip}) we obtain
\begin{eqnarray}
\frac{\partial \alpha_1}{\partial \theta_1}&=&
          \frac{\tilde{\kappa}+|\tilde{\gamma}|}{(1-\ephi)} 
          -2|\gamma|\frac{\theta_1^2}{(1-\ephi)^2\bar{\theta}^2}\,, \nonumber \\
\frac{\partial \alpha_2}{\partial \theta_2}&=&
          (\tilde{\kappa}+|\tilde{\gamma}|)(1-\ephi) 
          -2|\gamma|\frac{\theta_2^2(1-\ephi)^2}{\bar{\theta}^2}\,, \nonumber \\
\frac{\partial \alpha_1}{\partial \theta_2}&=&
\frac{\partial \alpha_2}{\partial \theta_1}=
          -2|\tilde{\gamma}|\frac{\theta_1\theta_2}{\bar{\theta}^2}\,. \nonumber
\end{eqnarray} 

Therefore the convergence is given by
\begin{eqnarray}
\kappa&=&A(\ephi)\tilde{\kappa}(\bar{\theta}) + B(\ephi,\theta_1,\theta_2)|\tilde{\gamma}(\bar{\theta})| \,,
\label{eq:kappa_elipphi}
\end{eqnarray}
where
\begin{eqnarray}
A(\ephi)&=&\frac{1+(1-\ephi)^2}{2(1-\ephi)}, \ \ \  
B(\ephi,\theta_1,\theta_2)=A(\ephi)-\frac{\theta_1^2+\theta_2^2(1-\ephi)^4}{\bar{\theta}^2(1-\ephi)^2} \,,\nonumber
\end{eqnarray}
and the shear components are given by 
\begin{eqnarray}
\gamma_1&=&C(\ephi)\tilde{\kappa}(\bar{\theta}) + D(\ephi,\theta_1,\theta_2)|\tilde{\gamma}(\bar{\theta})|\,, 
\label{eq:gamma1_elipphi}
\end{eqnarray}
where
\begin{eqnarray}
C(\ephi)&=&\frac{1-(1-\ephi)^2}{2(1-\ephi)}, \ \ \ 
D(\ephi,\theta_1,\theta_2)=C(\ephi)-\frac{\theta_1^2-\theta_2^2(1-\ephi)^4}{\bar{\theta}^2(1-\ephi)^2}\,, \nonumber
\end{eqnarray}
and
\begin{eqnarray}
\gamma_2&=&-\frac{2\theta_1\theta_2}{\bar{\theta}^2}|\tilde{\gamma}(\bar{\theta})|\,.
\label{eq:gamma2_elipphi}
\end{eqnarray}

Therefore, we can compute $\kappa$ and $\gamma_1$ and $\gamma_2$, 
at positions $\theta_1,\theta_2$ from the values of $\tilde{\kappa}$ 
and $|\tilde{\gamma}|$ at $\bar{\theta}$ from the spherical case.
The calculation of $|\gamma|$ and $\mu$ then follows. 

\section{Halo Mass Conversions}

Simulations of the SZ effect usually predict its flux decrement/increment
for different cluster 
mass definitions. 
Observationally it may be desirable to define
SZ halo masses at small apertures (large overdensities), since those can 
provide fluxes with larger signal-to-noise ratios. 
Moreover one expect smaller radii to enclose less flux from radio sources
and SMGs (though less SZ signal as well). 
However, as we show, smaller apertures also correspond to increased 
flux fluctuations from lensing.
We follow the mass conversion recipe outlined in 
\cite{Whi01} and \cite{HuKra03}, which we briefly review here for 
overdensities defined in terms of the {\it critical} density.

Converting different halo mass definitions requires assuming a prescription for 
the halo profile. 
Throughout, we have been assuming the NFW profile \citep{NFW97}
\begin{eqnarray}
\rho(r)=\frac{\rho_s}{(r/r_s)(1+r/r_s)^2}\,.
\end{eqnarray}

For this profile, the mass $M_h$ within radius $r_h$ enclosing 
an overdensity $\Delta_h$ with respect to the $critical$ density $\rho_c$ is
\begin{eqnarray}
M_h \equiv \frac{4\pi r_h^3}{3} \Delta_h \rho_c = 4 \pi\rho_s r_{h}^3f(r_s/r_h)\,,
\end{eqnarray}
where the first equality is a definition, the second results from the NFW profile
and 
\begin{eqnarray}
f(x)=x^3[\ln(1+x^{-1})-(1+x)^{-1}]\,.
\end{eqnarray}

For the usually employed virial mass $\Mvir$ the general equation above 
becomes
\begin{eqnarray}
\Mvir \equiv \frac{4\pi \rvir^3}{3} \Delta_c \rho_c = 4 \pi\rho_s \rvir^3f(1/c)\,,
\end{eqnarray}
where $c=\rvir/r_s$ is the halo concentration. From these equations we can relate
$M_h$ to $\Mvir$ as
\begin{eqnarray}
\frac{M_h}{\Mvir}&=&\frac{\Delta_h}{\Delta_c} \frac{1}{(cx)^3}\,, \\
x&=&\frac{r_s}{r_h}=f^{-1}\left(\frac{\Delta_h}{\Delta_c}f(1/c)\right)\,.
\end{eqnarray}

We employ the fitting formula for inversion of $x(f)=f^{-1}$ \citep{HuKra03}
\begin{eqnarray}
x(f)=\left[ a_1f^{2p}+\left(\frac{3}{4}\right)^2 \right]^{-1/2} + 2f\,,
\end{eqnarray}
where $p=a_2+a_3\ln f + a_4(\ln f)^2$ and 
$(a_1 , ... , a_4)=(0.5116, -0.4283, -3.13\times10^{-3}, -3.52\times10^{-5})$.
This fit is accurate to better than 1\% for the range of virial masses and 
concentrations used here.

\end{document}